\documentclass[a4,useAMS,usenatbib,usegraphicx]{mn2e}
\def\aapr{\ref@jnl{A\&A~Rev.}}

\usepackage{amsmath}
\include{epsf}
\usepackage{float}
\usepackage{color}
\usepackage{enumerate}

\usepackage{epsf}
\epsfclipon
\usepackage{amsmath}

\title[Modelling Auriga's Wheel]{Numerical modelling of Auriga's Wheel - a new ring galaxy}
\author[R. Smith et al]{R. Smith$^{1}$\thanks{E-mail:rsmith@astro-udec.cl}, R. R. Lane${^1}$, B. C. Conn${^2}$, M. Fellhauer${^1}$\\
$^{1}$Departamento de Astronomia, Universidad de Concepcion, Casilla 160-C, Concepcion, Chile\\
\noindent
$^{2}$Max Plank Institut f{\"{u}}r Astronomie, K{\"{o}}nigstuhl 17, 69117, Heidelberg, Germany}
\begin{document}

\date{Accepted to MNRAS 2012 March 9.  Received 2012 March 8; in original form 2011 January 13}

\pagerange{\pageref{firstpage}--\pageref{lastpage}} \pubyear{2011}

\maketitle

\label{firstpage}

\begin{abstract}
We model the formation of Auriga's Wheel - a recently discovered collisional ring galaxy. Auriga's Wheel has a number of interesting features including a bridge of stars linking the neighbouring elliptical to the ring galaxy, and evidence for components of expansion and rotation within the ring. Using $N$-body/SPH modelling, we study collisions between an elliptical galaxy and a late-type disk galaxy. A near direct collision, with a mildy inclined disk, is found to reasonably reproduce the general system morphology $\sim$50~Myr following the collision. The collision must have a relatively low velocity (initially $\sim$150~km~s$^{-1}$) in order to form the observed bridge, and simultaneously match the galaxies separation. Our best-match model suggests the total disk galaxy is $\sim$5 times more massive than the elliptical. We find that the velocity of expansion of the ring is sensitive to the mass of the elliptical, while insensitive to the encounter velocity. We evolve our simulation beyond the current epoch to study the future destiny of the galaxy pair. In our model, the nucleus moves further away from the plane of the ring in the direction of the stellar bridge. The nucleus eventually merges with the elliptical galaxy $\sim$100~Myr after the present time. The ring continues to expand for $\sim$200~Myr before collapsing back. The low initial relative velocity of the two galaxies will eventually result in a complete merger.

\end{abstract}


\section{Introduction}
For over half a century, astronomers have pondered the origin of ring galaxies. These rare and peculiar galaxies have a striking appearance characterised by a bright ring of gas and stars.

Some are now believed to form as the result of a close encounter between two galaxies and are called Collisional Ring Galaxies (CRGs). However a previous alternative hypothesis for their origin included a collision between a normal late-type disk galaxy and an intergalactic gas cloud (\citealp{Freeman1974}). This theory has been largely neglected; the required isolated gas clouds appear too rare to explain ring galaxies (\citealp{Appleton1996}), and new evidence favours a galaxy-galaxy collision origin in two out of three of Freeman and Vaucouleurs' prototype examples (e.g. two cores found in Arp144, \citealp{Joy1989}).

The success of modelling the origin of ring galaxies as a galaxy-galaxy encounter was first demonstrated by the numerical models of \cite{Lynds1976}. A companion galaxy moves along the axis of rotation of the disk, and eventually is close to the centre of the parent galaxy. This increases the gravitational attraction of the stars of the parent galaxy towards the centre of their disk. However, once the companion has passed, the stars spring back. This results in a well defined density ring that expands outwards from the centre of the disk. This study provided a conceptual model for the formation of the ring structure.

The rings of CRGs typically contain young, blue stars and a colour gradient is often present with increasingly red colours at smaller radii inside of the ring (see for example the multi-wavelength study of \citealp{Appleton1997}). This colour gradient is an expected outcome of an origin in which a companion galaxy passes through the disk of another galaxy. This generates an outwardly propagating density wave in the gas and stellar disks, and star formation is presumably enhanced within the higher density gas ring. Thus young, blue stars are found nearer to the ring. However, \cite{Gerber1994} note that the gas and stars may not respond alike to the perturbation. The gas density wave can lag behind that of the stellar density wave, in response to a large perturbation. If young stars are found in the gas ring, then this could potentially result in a steep reversal of the colour gradient outside of the gas ring. Such a reversal is seen in some ring galaxies (\citealp{Appleton1997}).

In the collisional formation scenario, the expanding ring may also contain some component of rotation - a relic of the angular momentum of the disk before collision. Observational studies, often using long slit optical spectroscopy, have confirmed signatures of combined rotation and expansion in ring galaxies (e.g. \citealp{Theys1976}, \citealp{Fosbury1977}, \citealp{Few1982}, \citealp{Jeske1986}, \citealp{Charmandaris1993}, \citealp{Conn2011}). Analytically, the expansion speed of the ring is predicted to scale with the mass of the companion galaxy (\citealp{Struck1990}).

\cite{Theys1976} constructed the first catalogue of ring galaxies, and provided a detailed morphological classification system. They noted that it is sometimes difficult to distinguish between rings formed by collisions and resonance-rings or pseudo-rings (\citealp{Vaucouleurs1959}; \citealp{Buta1994}) formed by tidal or bar resonances. However, they noted their ring galaxies predominantly had a companion within a projected distance of one or two ring diameters. 

\cite{Few1986} later compiled a larger catalogue and separated their sample into O-type (a central nucleus with a smooth and even outer ring), and P-type (with offcentre nucleus and clumpy ring). They found that their P-type sample had a statistically larger number of companions within two ring diameters than their control samples, whereas the O-type rings did not. They concluded that P-type rings were good candidates for CRGs. 

\cite{Few1986} also estimated the volume space density of ring galaxies as 5.4$\times 10^{-6}$~h$^3~$Mpc$^{-3}$ corresponding to an average of 1 ring galaxy in every sphere of radius $\sim35~$Mpc. This is approximately a factor of 10$^4$ less than the space density of average moderately luminous disk galaxies (\citealp{Appleton1996}). CRGs are, therefore, rare objects because their formation requires a galaxy-galaxy collision in a fairly narrow regime of parameter space. For example, \cite{Onghia2008} investigated the formation rate of CRGs in $\Lambda$CDM simulations, and consider that a CRG will form when a companion galaxy passes within 15$\%$ of the diameter of the disk of the parent galaxy (as established in \citealp{Lynds1976}), and that the mass ratio of the companion to the parent galaxy must be $>$1:10 (\citealp{HernquistWeil1993}, \citealp{Horellou2001}). They do not consider limits on inclination of the disk to the vector of motion of the companion galaxy, however \cite{Lynds1976} produce symmetric disks for $\theta<45^\circ$, and \cite{Ghosh2008} produce complete, albeit warped rings for $\theta>60^\circ$.

The impact parameter is the minimum distance between the parent and companion galaxy. The effect of varying the impact parameter (in this case measured in the direction of the plane of the disk) on the final ring morphology is clearly visible in Figure 5 of \cite{Toomre1978}. Increasing the impact parameter causes the ring to become increasingly lopsided and non-circular until, eventually, the disk forms a spiral structure rather than a ring. Increasing the impact parameter also shifts the nucleus further towards the denser side of the ring. Movement of the nucleus has been used to explain ring galaxies with no visible nucleus (RE galaxies). \cite{Gerber1992} modelled Arp147 as formed in such a collision and suggests that the nucleus is in fact far off-centre within the ring, and also behind the plane of the ring. As a result the nucleus is hidden behind the edge of the ring in his models. A final location of the nucleus behind the plane of the ring is also reported in \cite{Lynds1976} and \cite{Huang1988} (although see \citealp{Mapelli2011} whose model has the nucleus within the plane of the ring).

As the companion galaxy is typically less massive than the parent galaxy with which it collides, it is worth considering the impact on the companion galaxy too. `Mushroom galaxies' appear to be examples where the companion has been heavily disrupted by the encounter, and stretched out into the stalk of  a mushroom, while the ring galaxy composes the head of the mushroom. The archetypal example is referred to as `the Sacred Mushroom', and was first noted in the catalogue of \cite{Arp1987}. Observations combined with simulations suggest the ring/mushroom head may be a rare example of a gasless ring formed from the disk of a parent galaxy that, pre-collision, had an S0 type morphology, and that the stalk of the mushroom is well reproduced by the stretching out of the companion galaxy (\citealp{Wallin1994}).

Streams that link a ring galaxy to a companion galaxy, provide very strong evidence for the collisional origin of a ring. Indeed the Cartwheel galaxy, the archetype example of CRGs, appears to have a HI streamer linking the ring to a potential companion (\citealp{Higdon1993}). However, the presence (or lack of) a stellar or gaseous bridge likely depends on the velocity of encounter, as does the future evolution of the ring galaxy. \cite{Mapelli2008} examine the future evolution of ring galaxies formed in high velocity encounters (initial relative velocity of 900~km~s$^{-1}$). The high initial velocity forbids a merger, and the rings expand to increasingly large diameter and low surface brightness. In this manner ring galaxies formed in high velocity encounters may be the progenitors of giant low surface brightness galaxies.

\begin{figure}
  \centering \epsfxsize=6.5cm \epsfysize=6.0cm
  \epsffile{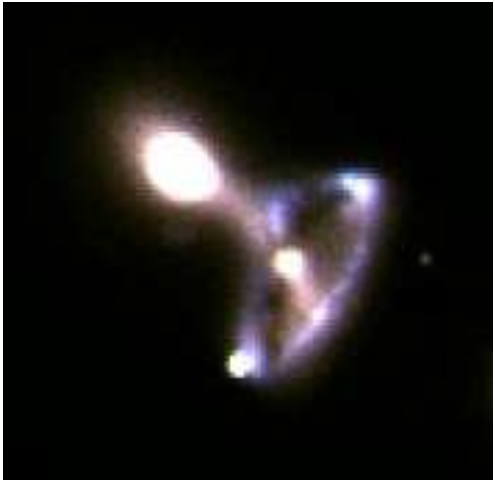}
  \caption{Subaru SUPRIME-CAM false-colour composite of Auriga's Wheel formed with data from a single $g'$ and $r'$ image. Box size is approximately 50 kpc (23 arc seconds). Pixel scale is 0.2 arc seconds.}
\label{rgb}
\end{figure}

In \citealp{Conn2011} (hereafter C11), a new ring galaxy was reported dubbed `Auriga's Wheel'. Optical imaging with Subaru SUPRIME-CAM reveals a close pairing ($\sim13~$kpc separation) between the ring galaxy and an elliptical companion. A false-colour composite of Auriga's Wheel is shown in Figure \ref{rgb}. A stellar bridge links the ring and elliptical, leaving little doubt about the collisional origin of the galaxy, nor the companion with which it collided. The ring is clearly P-type with a clumpy morphology and apparently mildly offcentre nucleus. The ring is blue ($g'$-$r'\sim0.7$) in comparison to the nucleus ($g'$-$r'\sim1.0$). C11 also presents long-slit optical spectroscopy (Gemini/GMOS-N) across regions of the ring. Under the assumption of a circular ring, the ring appears to be simultaneously expanding (at $\sim200~$km~s$^{-1}$), and rotating (at $\sim60~$km~s$^{-1}$).

We model the formation of the Auriga's Wheel in a low velocity collision between a massive late-type disk galaxy (with bulge) and the companion elliptical galaxy. The numerical code used and the initial set up of the galaxies is described in Section 2, the formation and time-scales of our best-match model are shown in Section 3, the model is compared to Auriga's Wheel in Section 4, a short parameter study is conducted in Section 5, modelling of the future evolution of Auriga's Wheel is shown in Section 6, and finally the discussion and conclusions are in Section 7.

\section{Setup}
\subsection{The Code}
In this study we make use of `gf' (\citealp{Williams1998},\citealp{Williams2001}), which is a Tree code-SPH algorithm that operates primarily using the techniques described in \cite{Hernquist1989}. While the Tree code allows for rapid calculation of gravitational accelerations, the SPH code allows us to include a gas component to our disk galaxy models. In all simulations, the gravitational softening length, $\epsilon$, is fixed for all particles at a value of 100 pc. Gravitational accelerations are evaluated to quadrupole order, using a tree code opening angle of $\theta_{\rm{c}}$=0.7. A second order individual particle time-step scheme was utilised to improve efficiency following the methodology of \cite{Hernquist1989}. Each particle was assigned a time-step that is a power of two division of the simulation block time-step, with a minimum time-step of $\sim$5.0 yr. Assignment of time-steps for collisionless particles is controlled by the criteria of \cite{Katz1991}, whereas SPH particle time-steps are assigned using the minimum of the gravitational time-step and the SPH Courant conditions with a Courant constant, $C$ = 0.1 (\citealp{Hernquist1989}). As discussed in \cite{Williams2004}, the kernel radius $h$ of each SPH particle was allowed to vary such that at all times it maintains between 30 and 40 neighbours within 2$h$. In order to realistically simulate shocks within the SPH model, the artificial viscosity prescription of \cite{Gingold1983} is used with viscosity parameters $(\alpha,\beta)$ = (1,2). The equation of state for the gas component of the galaxies is isothermal with a bulk velocity dispersion of 10.0 km\,s$^{-1}$, in agreement with the measured velocity dispersion of molecular clouds in the local interstellar medium (\citealp{Stark1989}), and the observed HI velocity dispersion within a radius containing significant star formation in late-type disks (\citealp{Tamburro2009}). By choosing an isothermal equation of state, we are intrinsically assuming that stellar feedback processes are balanced by radiative cooling producing a constant velocity dispersion. Such a model can be considered an approximation to a detailed description of a multiphase ISM that is unresolved in our models. Within real galaxy disks and collisional rings, gas is sufficiently dense that cooling time-scales are very short, effectively maintaining the gas in a near isothermal state (\citealp{Hernquist1993b}). An isothermal equation of state has been used previously in simulations of galaxy dynamics, galaxy mergers, and collisional ring galaxy formation (\citealp{Theis1993}, \cite{Hernquist1993b}, \citealp{Mihos1994II}, \citealp{Englmaier1997}). We do not include star formation in the models.

Our best-match model is realised in a near direct collision between a massive spiral galaxy, and a less massive elliptical galaxy. We arrive at our best-match model using a substantial amount of trial and error in terms of the initial galaxy properties, and their relative dynamics. We do not provide specific details of this process as we feel this is less scientifically useful than our following approach. Instead we first concentrate on the properties of our best-match model in Sections \ref{Elpgalmodel}--\ref{matchobs}. We shall then demonstrate the sensitivity and dependencies of our best-match model on a number of key parameters in Section \ref{paramstudy}.

\subsection{Elliptical galaxy model}
\label{Elpgalmodel}
We model the elliptical galaxy as a Hernquist sphere (\citealp{hernquist1990}). A Hernquist sphere has a density distribution following

\begin{equation}
\rho(r) = \frac{M_{\rm{h}}}{2\pi}\frac{r_{\rm{h}}}{r}\frac{1}{(r+r_{\rm{h}})^3}
\label{herneqn}
\end{equation}
where $M_{\rm{h}}$ is the total mass of the Hernquist sphere, $r_{\rm{h}}$ is the Hernquist scalelength, and $r$ is radius. 

We assume an isothermal velocity dispersion for the particles that make up the elliptical galaxy. Thus the particle velocities can be calculated from the distribution function (e.g. Equation 20 in \citealp{hernquist1990}).

The properties of the observed elliptical may approximately match the properties of the pre-collision elliptical if the interaction has not modified it significantly. We will later show in Section \ref{ELPmorph} that this is the case. Therefore our choice of parameter values for the elliptical is partially motivated by the observations, and partially motivated by the results of the simulations.

We assume that the total mass of the elliptical is 2$\times$10$^{11}$~M$_\odot$. This intrinsically includes a component of dark matter. The line-of-sight velocity dispersion for a Hernquist sphere of this mass peaks close to the galaxy centre at $\sim$230~km~s$^{-1}$ - in reasonable agreement with the observed central velocity dispersion of the elliptical (C11). If we assume a dynamical mass to light ratio of 2, we have a total stellar mass of 1$\times$10$^{11}$ $M_\odot$. This is then in reasonable agreement with the stellar mass of the elliptical in C11 which is derived to be 1.1$\times$10$^{11}$~M$_\odot$, based on its colour and luminosity.

We choose $r_{\rm{h}} = 1.5$~kpc. This ensures our effective radius matches that of the observed elliptical if the dark matter to star mass ratio is constant throughout the elliptical galaxy. The density distribution is truncated at 30~kpc corresponding to 20 scale-lengths. In practice we find that stability is not compromised if we truncate the density distribution at radii greater than 10 $r_{\rm{h}}$.

\subsection{Disk galaxy model}
Our spiral galaxy model consists of 4 components; an NFW dark matter halo (\citealp*{Navarro1996}), an exponential disk of gas and one of stars, and a stellar bulge. 

\subsubsection{The dark matter halo}
The dark matter halos of the disk galaxy model has an NFW density profile. The NFW profile has the form:
\begin{equation}
\rho(r) = \frac{\rho_0}{(\frac{r}{r_{\rm{s}}})(1+\frac{r}{r_{\rm{s}}})^2}
\label{NFWdensprof}
\end{equation}
\noindent where $\rho_0$ is the central density, and $r_{\rm{s}}$ is a characteristic radial scale-length. The profile is truncated at the Virial radius, $r_{\rm{200}} = r_{\rm{s}} c$. Here $c$ is the concentration  parameter (\citealp{Lokas2001}). $c$ is found to have a range of values in cosmological simulations, however there is a general trend for higher values in less massive systems with some scatter - see Figure 8 in \cite{Navarro1996}.

Positions and velocities are assigned to the dark matter particles using the publically available algorithm {\it{mkhalo}} from the {\sc{nemo}} repository (\citealp{McMillan2007b}). Dark matter halos produced in this manner are evolved in isolation for 2.5 Gyr to test stability, and are found to be highly stable.

Our best-match model for has a dark matter halo mass of $10^{12}$M$_\odot$, consisting of 100,000 dark matter particles, with a concentration $c = 18$. The Virial radius is $r_{\rm{200}} = 202$~kpc.

\subsubsection{The bulge}
We model the stellar bulge of the spiral galaxy as a Hernquist sphere (i.e. see Eqn. \ref{herneqn}). As with the elliptical galaxy, it is radially truncated at a cut-off radius. Particle velocities are assigned using the Jean's equation for an isotropic dispersion supported system.

The bulge of the disk galaxy has a mass of 2$\times10^{10}~M_\odot$, a Hernquist scalelength $r_{\rm{h}} = 0.4~$kpc, and a cut-off radius of 30~kpc. The bulge is formed from 20,000 star particles. 

\subsubsection{The disk}
The disk of the spiral galaxy model consists of a stellar and gas disk superimposed. The stellar and gas disk both have an exponential form
\begin{equation}
\label{expdisk}
\Sigma(R) = \Sigma_0 {\rm{exp}} (R/R_{\rm{d}})
\end{equation}

\noindent
where $\Sigma$ is the surface density, $\Sigma_{\rm{0}}$ is central surface density, $R$ is radius within the disk, and $R_{\rm{d}}$ is the scale-length of the disk.

The size of the stellar scale-length of the disk is chosen following the recipe of \cite{Mo1998}. Here, the disk mass is a fixed fraction, $m_{\rm{d}}$ of the halo mass. Additionally we must choose the spin parameter $\lambda$. In the Mo et al. recipe, this choice of parameters fully defines the scale-length of the stellar disk, $R_{\rm{d}}$. For simplicity we assume the scale-length of the stellar disk and gas disk are equal.

A radially varying velocity dispersion is chosen that ensures the disk is Toomre stable (\citealp{Toomre1964}) at all radii. In practice, a Toomre parameter of $Q>1.5$ is required throughout the stellar disk to ensure stability. A full description of the procedures followed to set up the galaxy disk can be found in \cite{Smith2010a}.

\begin{figure*}
\includegraphics[scale=0.75]{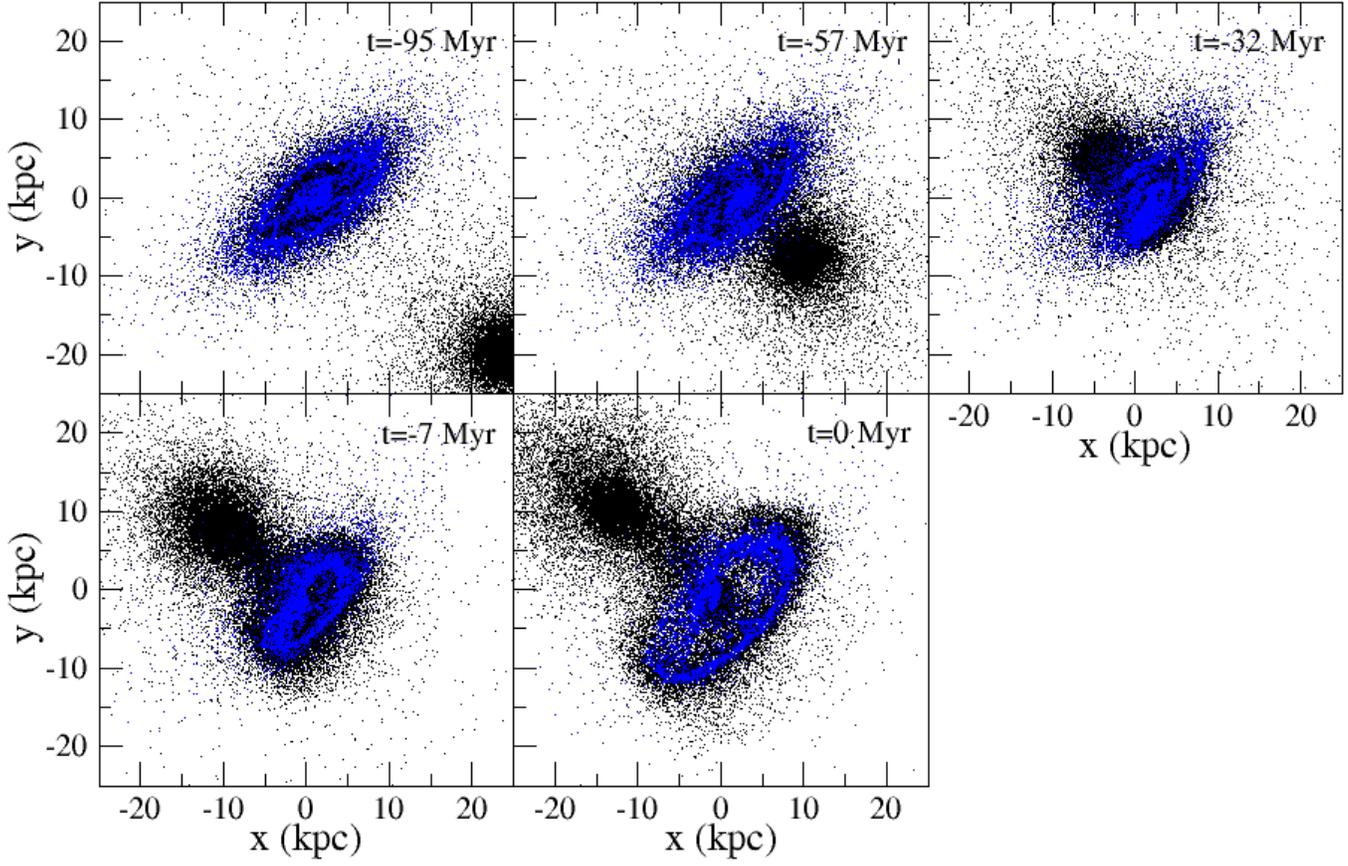}
\caption{Formation of the ring galaxy - snapshots taken of our best-match model from t = $-$95~Myr (upper-left panel) until t = 0~Myr (lower-left panel) corresponding to our current view of Auriga's Wheel. Star particles are black points, and gas particles are blue points.}
\label{mainfig}
\end{figure*}

Our best-match model has $m_{\rm{d}}$ = 0.05, therefore has a total disk mass of $5.0 \times 10^{10}$M$_\odot$, and is formed from 40,000 star particles and 50,000 gas particles. We arbitrarily choose the gas to stellar mass ratio of the disk to be 20:80 - at the upper limit of what is observed for z=0 giant late type disk galaxies (\citealp{Gavazzi2008}). We choose the spin parameter of our best-match model as $\lambda$=0.05 resulting in a stellar and gas disk scale-length of $R_{\rm{d}} = 3.0$ kpc.

\subsection{Initial location and dynamics of the galaxy-pair}
\label{icsposvel}
The centre of the elliptical model is always initially located at the origin in our simulation space, with no net velocity. We then place the center of the spiral at ($x$,$y$,$z$) = (68.4, 0.0, 187.9)~kpc such that the separation between the centres of the two galaxies is 202~kpc (the virial radius of the spiral).

We give the spiral an initial velocity of 150~km~s$^{-1}$ in the direction of the elliptical (($v_x$,$v_y$,$v_z$) = ($-$51.3, 0.0, $-$141.0)~km~s$^{-1}$). We also give the disk a 20$^\circ$ inclination in the x-z plane with respect to a vector joining the galaxy centres.

We offset the position of the spiral by 2.1~kpc (($x$,$y$,$z$)$_{\rm{final}}$ = (69.9, 1.5, 187.9)~kpc) to create a mildy off-centre collision, which we will later demonstrate to be of importance for reproducing the ring galaxy's morphology.

\section{Time-scales for formation of our best-match model}
Having described the initial location and dynamics, we now evolve the model forward toward the formation of the ring galaxy. The morphological evolution of the two galaxies can be seen in Fig. \ref{mainfig}. Panels present different snapshots of the stellar components (black) and gas component (blue), evolving in time. The time of the snapshot is indicated in Myr in the upper-right corner of each panel, where t = 0.0~Myr is the current appearance. 

\begin{figure*}
\begin{center}$
\begin{array}{ccc}
\includegraphics[width=6.0cm]{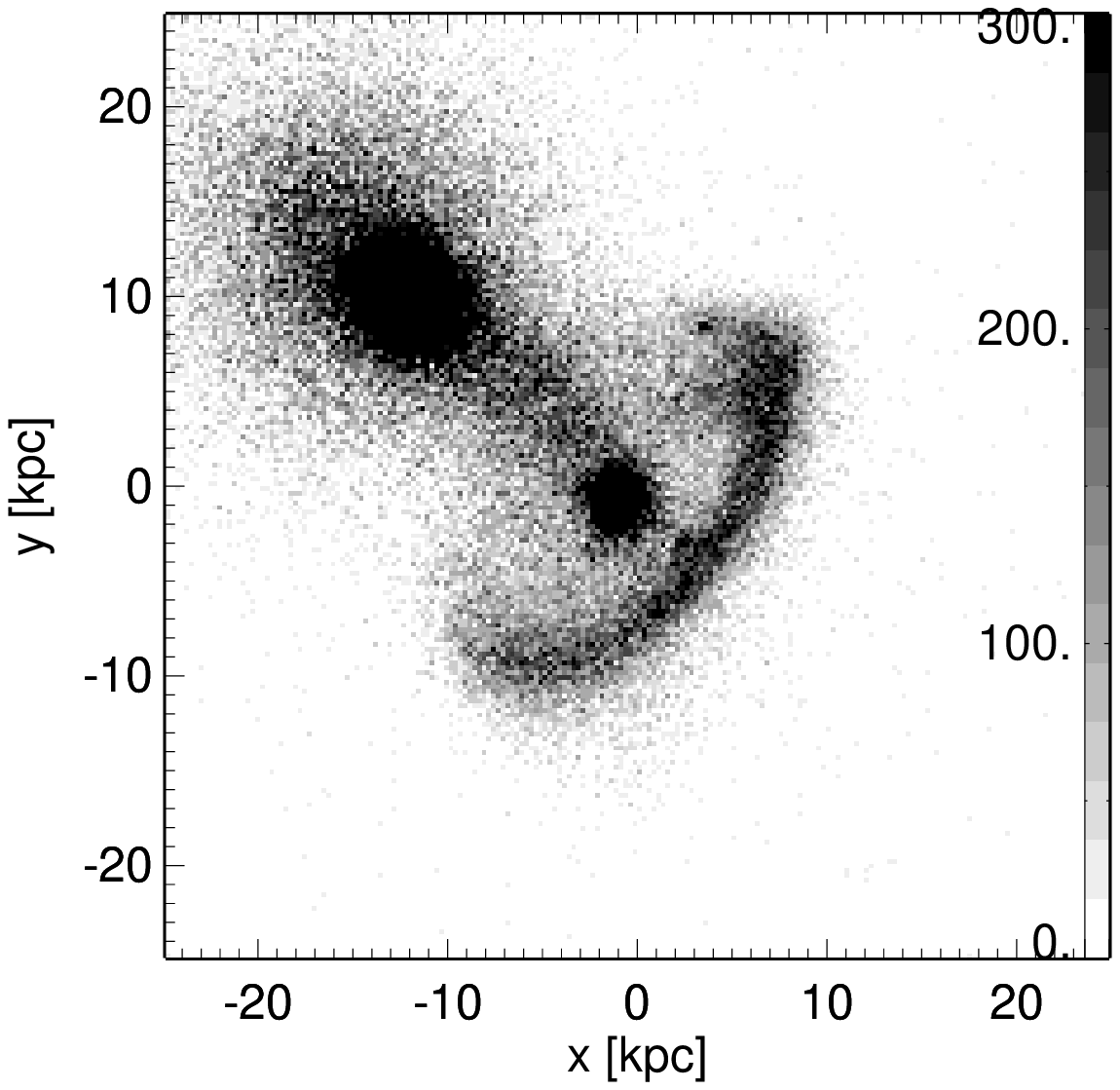} &
\includegraphics[width=6.0cm]{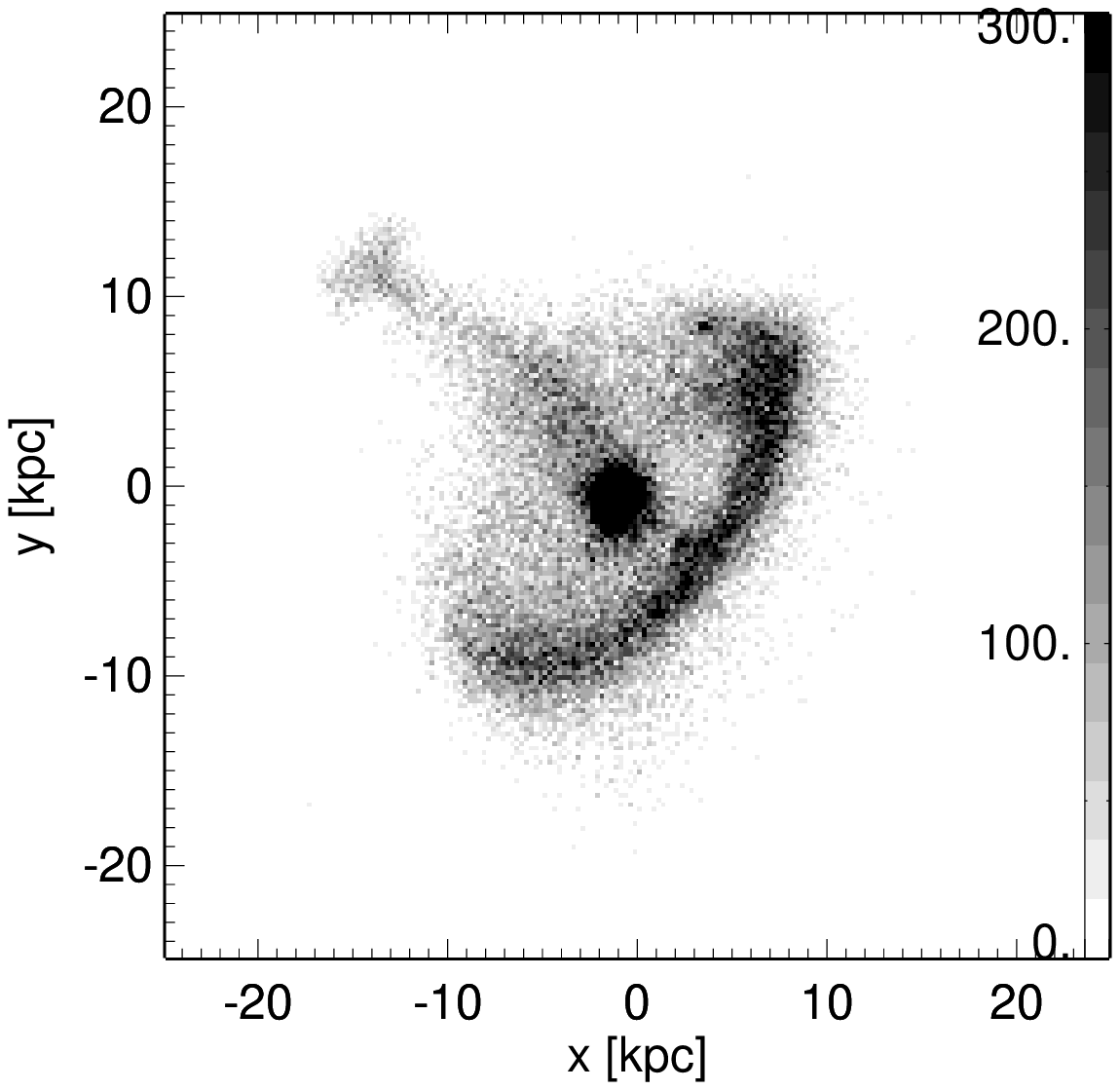} &
\includegraphics[width=6.0cm]{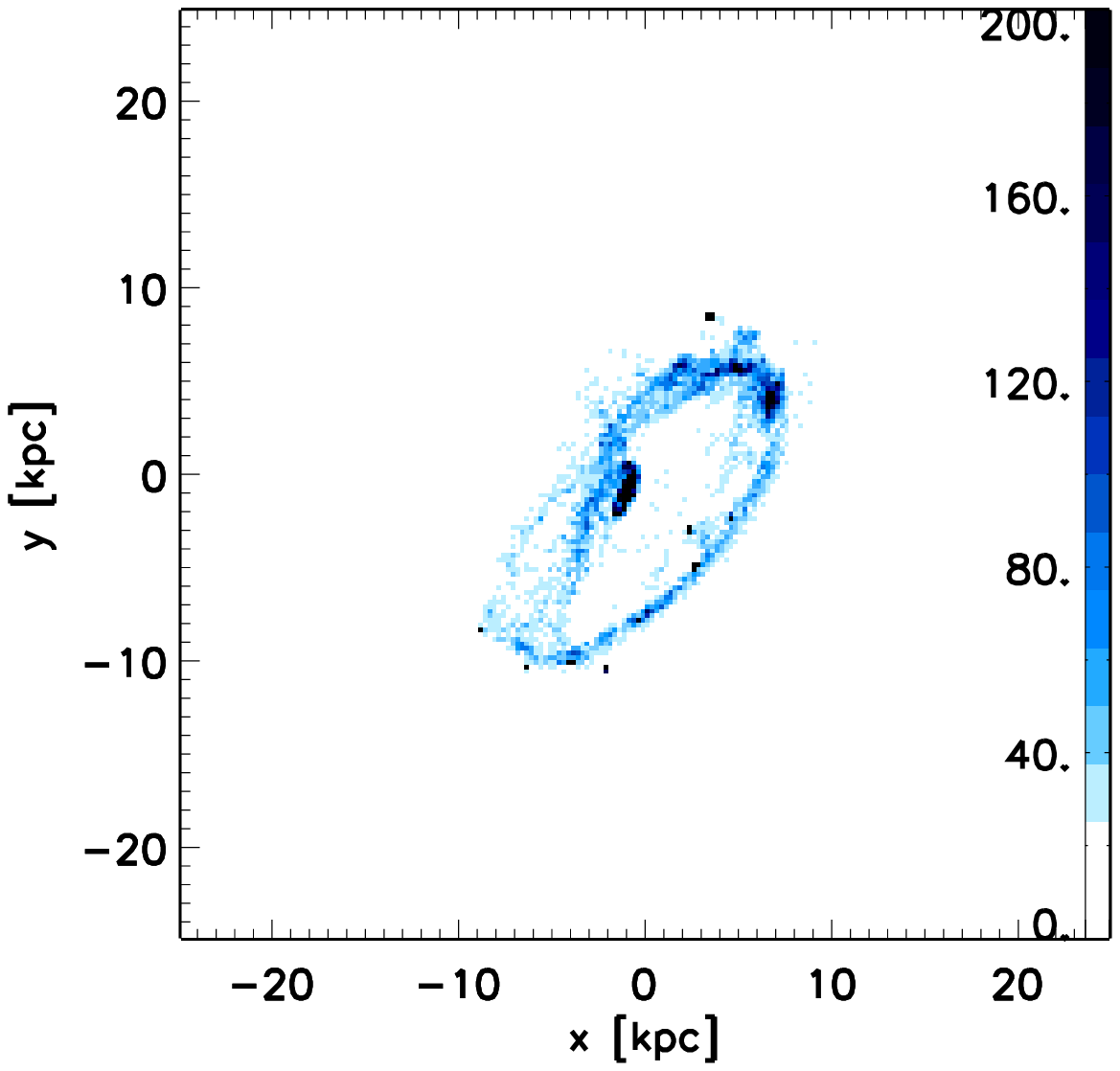} \\  
\end{array}$
\end{center}
\caption{Surface density plots of the best-match model; (left) stars from the original spiral and elliptical, (centre) neglecting the elliptical, and (right) the gas distribution. Colour bar labels in units of M$_\odot$~kpc$^{-2}$. The left and centre panels trace the older, fainter and redder disk stars that constitute the majority of the mass of the ring. However new stars are expected to form predominantly at the peaks of the gas density distribution. Hence the gas distribution in the right panel likely mimics the distribution of young, bright and blue ring stars.}
\label{nowfig}
\end{figure*}

Note that the upper left panel (t =$-95$~Myr) does not represent our initial conditions which would be found at t = $-795$~Myr. Thus from initial conditions to the formation of the current day ring galaxy involves almost 0.8~Gyr of evolution. However, only $\sim$40~Myr separates the instant at which the galaxies collide from the current day ring galaxy. Our best-match model therefore suggests that we are catching the ring galaxy very briefly after the galaxy-galaxy collision (less than 50~Myr). A small ring forms within the centre of the spiral disk at t = $-40$~Myr. This ring expands, sweeping up the surrounding outer disk gas, but also contains disk stars. At t = 0~Myr, the diameter of the ring is roughly the same size as the original gas disk.

In our models, the potential well of the elliptical galaxy acts to compress the inner dark matter, stars, and gas of the spiral galaxy. The collisionless components of the spiral galaxy model (the dark matter and stars) respond adiabatically to this compression. They are thus pressurised by the interaction. 

When the elliptical passes beyond the disk model's inner regions, the pressure is then released suddenly. The result is a density wave that propagates radially outwards from the galaxy centre - an expanding shell. The observable component of this density wave takes the form of a ring that expands radially outwards from the disk centre. The ring predominantly consists of the high angular momentum material from the original spiral disk. It therefore still contains an appreciable component of rotation, in order to conserve the angular momentum of the original disk.

However, the compression of the inner spiral by the elliptical's potential well can convert some initially rotationally dominated material to become dispersion dominated. We will show that this preferentially occurs to low angular momentum material located close to the centre of the disk. The result is a dispersion dominated nucleus found near the centre of the expanding and rotating ring.

\section{Reproducing the properties of Auriga's Wheel}
\label{matchobs}
In order to describe our best-match model, it is useful to first reiterate the key properties of Auriga's Wheel that we are attempting to match with our model. In this section, we list and enumerate these key properties. For ease of reference, we then ensure that the subsection which relates to a specific key property is numbered accordingly. We shall also refer to numbering system in the text. In Fig. \ref{rgb} we present a pseudo-color image of Auriga's wheel from Figure 2 of C11. For a comprehensive description of the Auriga's Wheel, please see C11.

\begin{enumerate}[(1)]
\item The elliptical and ring galaxy have a projected separation of $\sim$13~kpc. 
\item A stellar bridge links the elliptical and the ring galaxy.
\item The ring galaxy consists of an approximately circular ring of diameter $\sim$20~kpc, and an inner nucleus. The ring itself is significantly more blue ($g'$-$r'\sim$0.7) than the red ($g'$-$r'\sim$1.0) nucleus.
\item The stellar mass within the red nucleus is estimated to be $\sim$2.0$\times10^{10}$M$_\odot$ based on its colour and luminosity. There is an active galactic nucleus within the nucleus.
\item The stellar mass of the elliptical is estimated to be $\sim$1.1$\times10^{11}$M$_\odot$ based on its colour and luminosity. It has a central velocity dispersion of 230~km~s$^{-1}$, and the radial profile of its surface brightness, ellipticity, position-angle, and A4 parameter are well defined in Figure 3 of C11.
\item The relative velocity of the elliptical and ring galaxy down our line-of-sight is close to zero within errors ($\sim \pm$30km~s$^{-1}$).
\item The ring has a line-of-sight velocity of 50~km~s$^{-1}$ and 170~km~s$^{-1}$ away from us at the upper-right and lower right corner of the ring respectively. Assuming a circular ring with an inclination of 60$^\circ$ away from face-on, the ring has a component of radial expansion of magnitude $\sim$200~km~s$^{-1}$, and a component of rotation of magnitude $\sim$60~km~s$^{-1}$. 
\end{enumerate}

\subsection{Separation between the galaxy pair}
We now consider the t = 0~Myr results of our best-match model, corresponding to our current view of Auriga's Wheel. We first refer the reader to Fig. \ref{nowfig}. For clarity we have separated the snap shot at t = 0~Myr into components; left is the stars of the elliptical and disk galaxy, centre is the stars of the disk galaxy alone, and right is the gas of the disk galaxy. The colour bar (right of each panel) indicates the surface density of each component in M$_\odot$~kpc$^{-2}$. Between the nucleus of the ring and the elliptical galaxy we estimate a separation of 14~kpc, in reasonable agreement with point (1).

\subsection{The stellar bridge}
With the elliptical removed (central panel of Fig. \ref{nowfig}), the stellar bridge can be clearly seen and is fairly collimated in agreement with the observed bridge. We measure the total mass of the stream to be $\sim$5.5$\times 10^{9}$~M$_\odot$ and the majority of its length consists almost purely of stars (see central panel versus right panel of Fig. \ref{nowfig}). The source of the stars is approximately two-thirds disk stars and one-third bulge stars from the spiral. Very few stars from the elliptical are found in the stream. 

A small fraction of the gas ($<5\%$) is beginning to enter the stream at t=0~Myr, but it appears to be delayed in its flow along the stream in comparison to the stars.

The unusual shape of the end of the bridge forms as stars in the stream fall into the potential well of the elliptical. As they pass through the elliptical's centre, they are scattered creating a fan-like structure. However the bridge has low surface brightness in comparison to the centre of the elliptical. For an assumed stellar mass-to-light ratio of $\sim$7 in the $r'$-band, the surface brightness is approximately $\sim$24~mag~arcsec$^{-2}$.

The stream and fan shape is a direct prediction of the model, and it would be interesting to test the model against the observed stream's shape. However in the image, the elliptical is significantly higher surface brightness than the stream in the area where the fan shape is predicted to occur. We have attempted to subtract away the elliptical galaxy using fits to the light distribution based on a double-sersic, using iraf's {\sc{b-model}}, and using un-sharp masking. However the results have proven inconclusive - such a feature is visible in the `elliptical-subtracted' images when using a double-sersic, but is not clearly visible using the {\sc{b-model}} or unsharp masking approach. Unfortunately, the feature is found at the very core of the elliptical galaxy, where it is difficult to fit the elliptical's light distribution sufficiently accurately to conduct this test.

\subsection{The ring}
\label{ringmorph}
We draw the reader's attention to the difference in the morphological appearance of the stellar and gas component of the spiral galaxy following the galaxy collision (compare centre and right panel of Fig. \ref{nowfig}). The ring in the gas distribution is sharply defined in comparison to the more fuzzy distribution of stars. This difference is due to the dissipational nature of the gas in comparison to the collisionless nature of the stars. As a result the gas does not `bounce-back' in the same manner as the stars once the elliptical has passed. Qualitatively this behaviour is similar to that described in \cite{Gerber1994} in response to a massive perturbation.
 
We note that we have not included a star formation recipe within our simulation due to the simplistic, first order treatment of the gas physics within our code. However, it is highly likely that new young stars would be found close to the peaks of the density distribution of the gas. If so, the stellar distribution in the central panel of Fig. \ref{nowfig} represents the distribution of the older, redder, pre-collision stars of spiral galaxy. Whereas the right panel is likely to be a reasonable representation for the young, blue stars formed post collision. If so, a steep reversal of the colour gradient as we move beyond the ring might be expected -  from the ring of young blue stars in the gas ring to the ring of redder stars that were originally in the pre-collision disk.

We repeat our simulation varying our choice of assumed gas sound speed from 7.5-12.5~km~s$^{-1}$. Enhanced star formation and the accompanying feedback could potentially raise the gas sound speed beyond the 10~km~s$^{-1}$ that we have assumed, and this could occur predominately where young stars form and, as such, unevenly. By varying this parameter we can test the sensitivity of the final gas distribution (and thus the young stars) to our choice. We find that, independent of our choice of gas sound speed, the gas distributions have the same general shape, and ring size (although the exact location of clumps along the ring can differ somewhat). The expansion velocity of the ring is sufficiently high that it will be much larger than any realistic choice of sound speed within the gas. Therefore it always results in a shocked ring of expanding gas, and thus the general morphology and dynamics are largely unaffected by the exact choice of gas sound speed.

We measure gas densities within the outer disk only (at radius $>$2.5~kpc, thus excluding the gas within the nucleus). We find that if we assume a Schmidt-law type star formation ($\propto \rho_{\rm{gas}}^{1.5}$), star formation rates in the outer disk have doubled post-collision. Although only doubled, the star formation is limited to the ring only (excluding the nucleus), whereas pre-collision it was spread throughout the disk. Thus star formation per unit area is increased by much more than a factor of two, as it has become concentrated in the ring.

An additional result of the dissipational response of the gas to the collision is that the stars in the ring are distributed to larger radii than than the more compact ring in the gas distribution. As with the flow of gas and stars along the bridge, the gas appears to be slightly delayed in its response if compared to the stars. This behaviour is also noted by \cite{Gerber1994} in response to a large perturbation Although it is difficult to confirm this for the observed ring due to the differing seeing conditions for the g´-image and r´-image (see Fig. \ref{rgb}), there is perhaps a hint that the redder, older stars extend beyond the young blue stars in the manner seen in the model. Similarly, a reversal in the colour gradient at radii beyond the bright ring has been observed in other ring galaxies (\citealp{Appleton1997}). 

The 20~kpc diameter of the ring quoted in C11, and repeated in point (3), is with reference to the bright, young, blue stars in the ring. We note that the right hand panel of Fig. \ref{nowfig} is in good agreement with this result if young stars indeed form at the peaks of the gas density distribution as might be expected. 

\subsection{The nucleus}
\label{nuclmorph}
The model clearly shows that gas has been driven into the nucleus of the ring galaxy (see right panel of Fig. \ref{nowfig}). The mass of gas driven to the centre in our model is $\sim45\%$ of the total disk gas of the spiral. We therefore might expect a considerable starburst to occur within the nucleus. If all the gas was consumed in the starburst, it would create 4.5$\times10^{9}$~M$_\odot$ new, young and blue stars. This qualitatively appears to be in contradiction with point (3) - the observed colour of the nucleus is noticeably more red than the ring itself.

\begin{figure}
  \centering \epsfxsize=8.5cm \epsfysize=5cm
  \epsffile{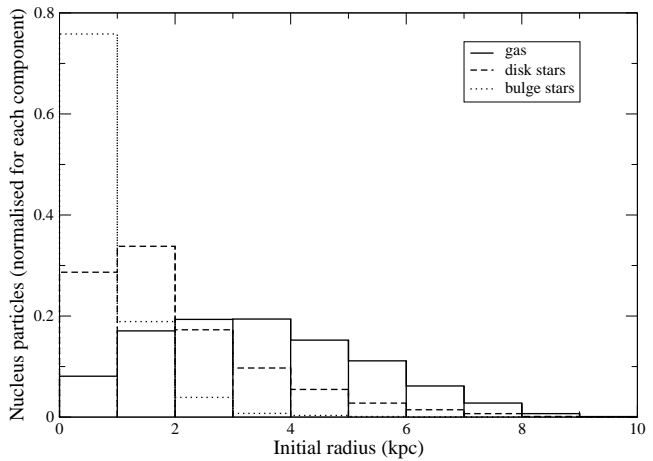}
  \caption{Particles that are found in the post-collision nucleus of the ring galaxy are re-traced to their radius in the pre-collision disk galaxy. Histogram of the initial radius of; gas (solid line), disk stars (dashed line) and bulge stars (dotted line) are compared. Nucleus stars originated from the inner regions of the disk galaxy. Nucleus gas originated from further out in the disk. Each separate component of the nucleus if normalised separately.}
\label{nucleusradius}
\end{figure}

However quantifying the colour of the nucleus as predicted by the model is challenging. The model suggests the nucleus would consist of a combination of old red bulge stars, old disk stars, and a young starbursting population. SSP modelling demonstrates that an actively starbursting population can alter its luminosity and colour on very short timescales, and we predict that the ring and nucleus were formed $<10~$Myr ago. Thus the stellar populations are in an extremely unpredictable regime, and furthermore may have their colours significantly effected by the dust that was likely to have existed in the original spiral. The same is true for the star-bursting population of stars within the ring. 

Excess central gas inflow has often been seen in CRG modelling (e.g. \citealp{Hernquist1993b}, \citealp{Horellou2001}, \citealp{Mapelli2008a}). We therefore try to understand the origin of the nucleus gas in the rest of this section.

We find the driving of material into the nucleus predominantly occurs to low angular momentum material (i.e. from the inner disk of the pre-collision disk galaxy where rotation velocities are low). To demonstrate this, we measure the properties of the particles within the model ring galaxy's nucleus and then trace back where these particles originated from. We consider all particles found within 2.5~kpc of the centre of the nucleus to be nucleus particles. The total stellar mass of the nucleus in the model is $\sim$1.5$\times 10^{10}$M$_\odot$. This is in reasonable agreement with the observed absolute magnitude. With $M_{r'}$ =$-19.05$, and $g'$-$r'$ = 1.04, we calculate the stellar mass-to light using Equation 1 of C11, and we find an upper limit to the stellar mass of 2.1$\times$10$^{10}$M$_\odot$ (i.e. point (4) of the key observed properties). If a starburst had indeed converted all gas that ends in the nucleus into stars, the total mass of the nucleus would be $\sim$1.9$\times 10^{10}$~M$_\odot$. Although such a starburst would likely result in our estimate for the stellar mass-to-light ratio to be highly uncertain.

Very few (0.4$~\%$) of the elliptical's stars are found in the nucleus, thus it predominantly consists of stars and gas from the spiral. 48.2$~\%$ of the original bulge of the spiral is found in the nucleus, compared to only 10.1$~\%$ of the original disk stars. The bulge is more centrally concentrated than the disk, and is dispersion supported, thereby containing a significant fraction of low angular momentum material. In fact, two-thirds of the mass of the nucleus is in the form of (pre-collision) bulge stars. Hence we find it difficult to form a sufficiently massive nucleus without a bulge component in the original spiral (as we shall demonstrate in Section \ref{nobulge}). Despite the fact that the gas and stellar disk have equal scalelength initially, 46.7$~\%$ of the spiral's disk gas ends up in the nucleus in comparison to the previously quoted 10.1$\%$ of the disk stars further highlighting the dissipational response of the gas during the compressional phase).

The original radius within the spiral, from which the various nucleus components originated (gas, disk stars, bulge stars) is presented in Fig. \ref{nucleusradius}. For bulge stars, $\sim$75$\%$ originated from within 1~kpc of the centre of the original spiral. For disk stars, $\sim$80$\%$ originated from within 3~kpc - one disk scale-length. Therefore the stellar component of the nucleus originated from the inner regions of the disk. However the original gas radii are much more extended. Only 45$\%$ is found at one disk scale-length, and $\sim80\%$ is found at 5~kpc. 

\begin{figure}
\centering
\includegraphics[angle=-90,scale=0.37]{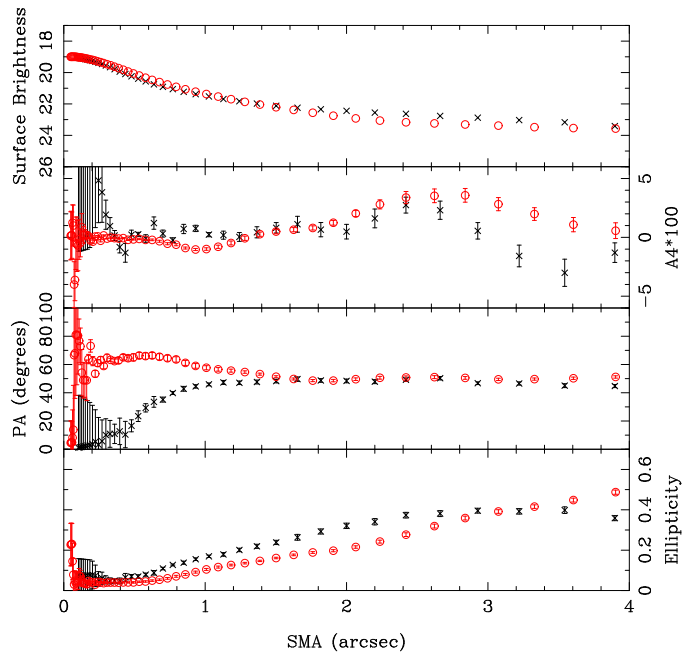}
\caption{Radial profiles of the elliptical galaxy, fitted using iraf's {\sc{ELLIPSE}} package. From upper to lower panel; ellipticity, position angle, surface brightness (in units of mag/arcsec$^2$) and 100$\times$A4 parameter. The x-axis is the semi-major axis (SMA) of the fitted ellipses in units of arc-seconds (1 arc-second = 2.2 kpc at the distance of Auriga's Wheel). Red symbols are the model and black symbols are the observed values.}
\label{radprofs}
\end{figure}

Therefore the quantity of gas driven to the nucleus centre will be dictated by the quantity of gas located in the inner $\sim5$~kpc of the original spiral disk and this will be sensitive to the gas surface density profile assumed. We have made the typical assumption of an exponential disk of gas for the original spiral disk. However, high resolution studies of the HI distribution in nearby giant spirals (e.g. the THINGS survey, \citealp{Portas2009}), demonstrate that the HI density distribution appears to have a more flat surface density distribution than that of an exponential disk. Furthermore it is well known that late-type spirals with significant bulge components, often show a HI hole in their inner regions. Our model spiral galaxy would indeed fall into this category, and so perhaps should have contained less, if any gas at all, in its inner disk. However, to completely quench a starburst would require a central HI hole of radius greater than $\sim$5~kpc.

It is worth noting that some gas inflow is required to fuel the AGN activity detected in the spectra of the ring galaxy's nucleus (i.e. point (4) of the key observed properties, and see Figure 4, C11). Furthermore complex feedback processes between the AGN and infalling gas could conceivably heat the gas preventing a starburst, and enabling the nucleus to maintain its red color.

\subsection{The elliptical morphology}
\label{ELPmorph}
In C11, radial profiles of the elliptical galaxy that neighbours Auriga's Wheel are presented including ellipticity, position angle, surface brightness, and the A4 parameter (point (5) of the key observed properties, \citealp{Bender1988}). In order to fairly compare between the observed radial profiles and the model, we produce model fits images. 

The model of the elliptical galaxy has a total mass of 2.0$\times10^{11}$~M$_\odot$, which is approximately half stars and half dark matter. If we derive a stellar mass-to-light ratio from the elliptical's colour (for a $g'$-$r'$ = 0.93, the r-band stellar mass-to-light ratio is 5.2), this mass of stars generates the observed luminosity of the elliptical (M$_{r'}$ = $-20.99$). The total mass of the elliptical is also sufficient to generate the observed central velocity dispersion of the elliptical (235~km~s$^{-1}$).

In many ellipticals, dark matter only begins to dominate their stellar dynamics in their outer radii (\citealp{Gerhard2001}). Therefore we allow the stellar fraction of the mass density to vary with radius. We choose the mass-density to be 85$\%$ stars at the elliptical centre, but dark matter begins to dominate the mass-density at a radius of $\sim 3~$kpc ($\sim 1.6 r_{\rm{eff}}$).  As we shall demonstrate, this provides a good match to the observed surface brightness profile.

Using these assumptions, we have now fully defined the luminosity and distribution of stars in the elliptical, and so can produce model fits images by projection of the three-dimensional distribution of the model into the two-dimensions of the image. 

\begin{figure}
\centering
\includegraphics[scale=0.35]{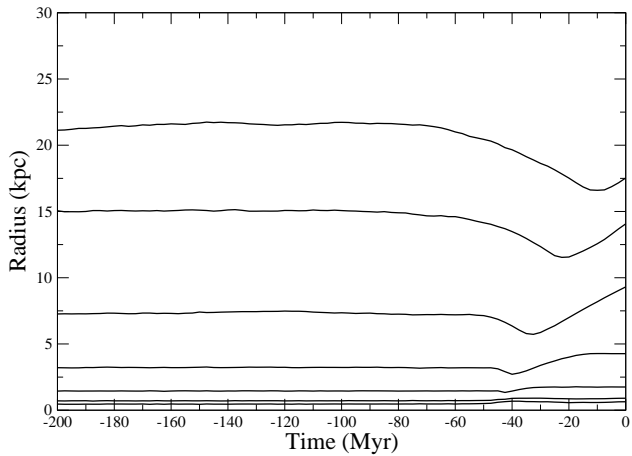}
\caption{Time evolution of the Lagrangian radii for the elliptical's mass distribution, from pre-collision (t = 200~Myr) until the current time (t = 0~Myr). From lower to upper respectively, Lagrangian radii contain 5, 10, 25, 50, 75, 90 and 95$\%$ of the total elliptical mass. The 25$\%$ (or less) Lagrangian radius  appears only mildy effected by the interaction, whereas the 75$\%$ (or more) Lagrangian radius outer is noticeably compressed, and reexpands.}
\label{elplagrangerad}
\end{figure}

We can then measure the model radial profiles directly from the fits images using the same software - iraf's {\sc{ELLIPSE}} routine. The results are shown in Fig. \ref{radprofs}. Radial profiles are shown out to a limiting semi-major axis (SMA) of 4.0 arc seconds (corresponding to $\sim8.8~$kpc) from the centre of the elliptical to avoid contamination by light from the ring.

Cross symbols (black) represent the observed profile, and circular symbols (red) represent the model profile. There is clearly a good agreement between the observed and modelled surface brightness profile (upper panel in Fig \ref{radprofs}).

Next we discuss the A4 parameter profile (second panel of Fig. \ref{radprofs}). Once more, within the errors we claim reasonable agreement between the model and Auriga's Wheel. As noted in C11, a switch from positive to negative values of the A4 parameter occurs as a result of the strong tidal encounter.

There is also excellent agreement with the observed position angle in the outer regions of the elliptical (third panel of Fig. \ref{radprofs}). However inwards of $\sim$1-2~arcsecs, the model and observed position angles diverge - with the model remaining approximately constant with radius, where as the observed profiles drops rapidly to close to zero in the $r'$-band. However it should be noted that the drop is much less in the $g'$-band (see Figure 3, \citealp{Conn2011}), bringing into question the real value of the position angle at the centre of the elliptical.

We also emphasise that this low position angle is measured where the galaxy is almost circular (lower panel of Fig. \ref{radprofs}), perhaps further reducing the significance of the apparent discrepancy. However, the ellipticity rises more slowly in the model, in comparison to the observations. 

One possible explanation for both of these failures is that, unlike our model elliptical, the original pre-collision elliptical was not spherical at all radii. Outside of the inner most contours (where the model and observations agree well), perhaps the pre-collision elliptical galaxy had mildly elliptical contours with a position angle that was close to zero in the inner regions.

However, following the galaxy-galaxy interaction, the outer regions of the elliptical were altered, whereas the inner regions remained largely unchanged. As a result, the outer regions have increasing ellipticity, a negative A4 parameter, and the direction of this elongation is along the axis of interaction of the two galaxies - a position angle of 60$^\circ$. In the inner regions, the elliptical remains close to spherical, but with a signature of its original position angle still remaining. 

Some support for this picture can be found in the evolution of the Lagrangian radii for the elliptical's stars, as presented in Fig. \ref{elplagrangerad}. The 25$\%$ Lagrangian radius is largely unaffected by the galaxy-galaxy interaction. However a more notable compression and re-expansion of the stars can be seen in the $75\%$ lagrangian radius. Thus the model predicts that the inner 2~kpc (corresponding to the 25$\%$ Lagrangian radius at t = 0~Myr) are largely unaffected. This corresponds closely to the radius at which the position-angle rises steeply to the outer values.

The lack of significant effect on the inner Lagrangian radii is perhaps reassuring for our choice of initial elliptical model, as we choose to match the initial effective radius to that of the current observed elliptical. If the elliptical had been heavily disturbed throughout then this would have been a poor decision.

In summary, our simple model of a spherical elliptical is probably a little too simple. The real pre-collision elliptical may have been mildly elliptical before the collision with a low position angle. It then had its outer regions altered by the collision.

\begin{figure}
  \centering \epsfxsize=8.5cm \epsfysize=6.0cm
  \epsffile{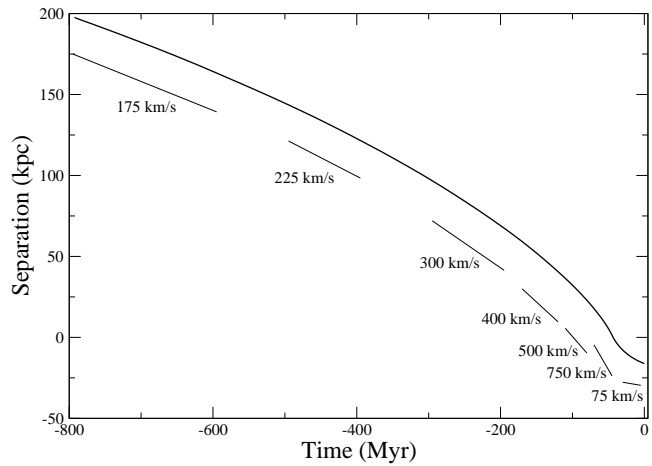}
  \caption{Time evolution of the relative distance between the elliptical model and spiral galaxy. We indicate how the gradient of the line corresponds to a velocity. Initially we set the relative velocity as low (150~km~s$^{-1}$). The mutual gravitational attraction accelerates the galaxies with respect to each other. At t = $-40$~Myr (corresponding to the instant that the elliptical and spiral are superimposed), the relative velocity briefly peaks at $\sim$750~km~s$^{-1}$. Following the collision, the relative velocity decreases rapidly to $<$75~km~s$^{-1}$ at t = 0~Myr, and continues to fall.}
\label{relpos}
\end{figure}

\subsection{Relative velocity of the galaxies}
In Fig. \ref{relpos}, we present the relative distance-separation of the spiral and elliptical and its evolution with time (thick curve). As discussed in Sect. \ref{icsposvel}, the spiral is initially positioned at $\sim$200~kpc from the elliptical, and given a 150~km~s$^{-1}$ velocity vector directed towards the centre of the elliptical. We indicate what the gradient of the line indicates in terms of their relative velocity (thin straight lines beside the thick curve). The relative velocity steadily rises due to the gravitational attraction of the spiral and the elliptical galaxy. 

The relative velocity peaks briefly at $\sim$750~km~s$^{-1}$ when the elliptical and disk of the spiral are superimposed (t=$-40~$Myr). However, dynamical friction slows down their relative velocity considerably. This is a natural consequence of the kinetic energy transfer from orbital kinetic energy of the elliptical to internal kinetic energy of the spiral. In the model, we estimate the t = 0~Myr velocity to be $<$75~km~s$^{-1}$. Although this is low, it is perhaps a little higher than seen in Auriga's Wheel even accounting for inclination (see point (6) of the key observed properties). However, we note the quoted errors on the relative velocity are lower limit estimates.

\begin{figure}
\begin{center}$
\begin{array}{cc}
\includegraphics[width=9.0cm]{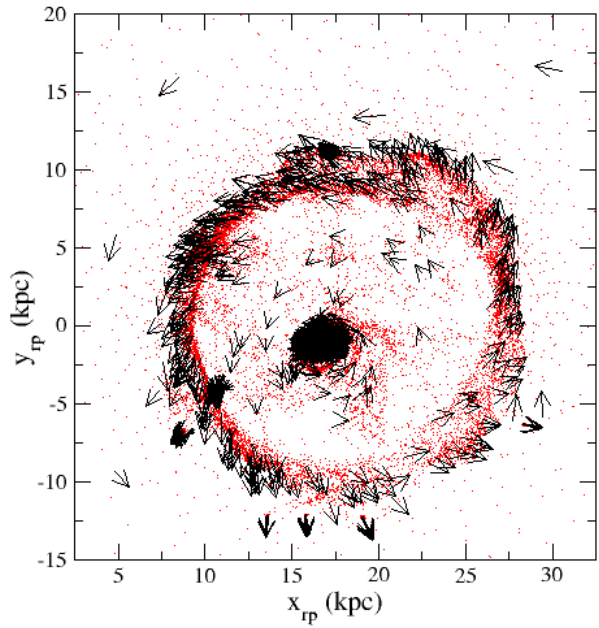} \\  
\end{array}$
\end{center}
\caption{The ring in the gas component, with velocity vectors. The ring is rotated so as we see it face-on (x and y coordinates have the subscript `rp' to denote that these are coordinates in the plane of the ring). A combination of rotation and expansion away from the ring centre can be seen, resulting in a twist in the velocity field away from radial expansion. Despite a mildly off-centre collision, the ring remains close to circular and expands at a roughly uniform velocity, however the nucleus is clearly off centre within the ring.}
\label{vxyring}
\end{figure}

\subsection{Ring dynamics}
\label{ringdyn}
We now focus on the dynamical properties of the ring itself at t=0~Myr. In Fig. \ref{vxyring} we present the gas particles as points (red) with super-imposed velocity vectors (black arrows). The gas ring is clearly close to circular despite the off-centre collision. Thus the assumption of an inclined circular ring, used to derive the expansion and rotation component of the ring in C11, is reasonable. This is also in agreement with \cite{Lynds1976} who find they can produce circular rings for impact parameters up to 15$\%$ of the disk outer diameter. Although in our case the impact parameter is only $\sim$5$\%$ of the disk at 5 exponential disk scalelengths.  However, the nucleus has not formed in the centre of the ring and is offset approximately 2-3~kpc. The nucleus has been kicked off centre by the mild off-centre nature of the collision. This behaviour is qualitatively similar to that seen in \cite{Toomre1978}. In the following section, we show that this is important for reproducing the observed location of the nucleus within the ring.

Additionally note that in general, the gas within the ring is expanding, however there is an additional element of rotation (anti-clockwise in the x-y plane presented). As expected, the ring maintains a signature of the rotation of the spiral from which it formed, although it expands radially outwards from the centre of the ring as a shockwave. At t = 0 Myr, we measure the ring expansion speed as $\sim$210~km~s$^{-1}$, in reasonable agreement with point (7) of the key observed properties.

\begin{figure}
  \centering \epsfxsize=8.5cm \epsfysize=8.5cm
  \epsffile{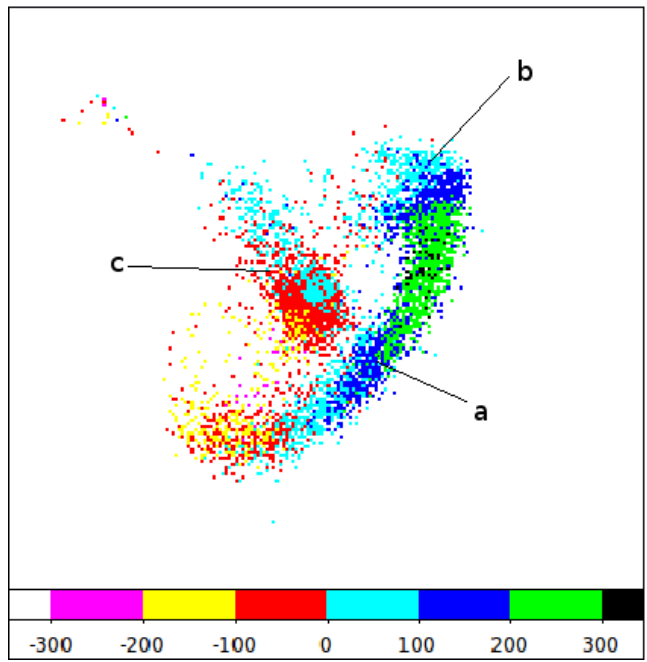}
  \caption{Colour coded line of sight velocities of the disk stars of the original spiral model. We label 3 positions on the ring corresponding to locations where spectroscopic data has provided line-of-sight velocity measurement for Auriga's Wheel. We find reasonable agreement at all three locations (see text for further details). Box size is 40 kpc.}
\label{losvels}
\end{figure}

In Fig. \ref{losvels}, we present the colour-coded, average velocity down our line-of-sight of the spiral disk stars with respect to the nucleus. Note that a positive velocity corresponds to receding stars (following the same convention as Table 4 in C11). Evidence for the combined components of expansion and rotation can be seen within the velocity field. 

For example, if the ring were purely expanding, we would expect to see the maximum receding velocity at position `a', and it would have magnitude equal to the expansion speed of the ring multiplied by the cosine of the angle of the plane of the disk with respect to our line-of-sight. The maximum approaching velocity would be found on the opposite side of the ring to position `a'. In a similar manner we would expect to see a line-of-sight velocity close to zero at position `b', and also on the opposite side of the ring.

\begin{figure}
\begin{center}
\includegraphics[scale=0.37]{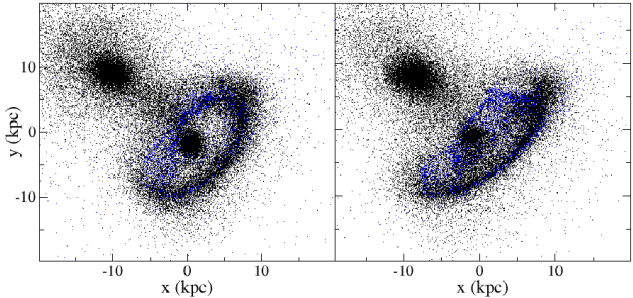} \\  
\end{center}
\caption{At t = 0 Myr, the distribution of stars (black points) and gas (blue points) in the best-match model (left), and in the high resolution re-run (right).}
\label{restest}
\end{figure}

However the ring is also rotating. This causes the velocity at position `b' to be non-zero. The component of rotation in Fig. \ref{losvels} is such that at `b' it rotates away from our line of sight (positive colour-coded values). On the opposite side of the ring from position `b' (the bottom-left edge of the ring) it rotates toward our line-of-sight (negative colour-coded values). 

Position `b' is special in that we can not see any of the ring's expansion component here, as it will all be directed perpendicular to our line-of-sight. Thus we only see the rotation component at position `b'. We measure this to be $\sim$50 $\pm$ 20~km~s$^{-1}$. The large errors reflect the considerable scatter between neighbouring pixel values. This is in agreement within errors with point (7) of the key observed properties of Auriga's Wheel after accounting for the ring inclination.

In a similar manner, at position `a' we expect all the component of rotation is directed perpendicular to our line of sight, so as we will only see the expansion component. This is measured to be $\sim$180 $\pm$ 20~km~s$^{-1}$, the large errors once more reflecting the noise in neighbouring pixel values. This is also in agreement with the observed value within errors (again see point (7)). 

Although this region of the ring is not observed, we note that the component of rotation causes the maximum receding velocity on the ring to be found a little above position `a' (on the ring between position `a' and `b'). The combination of the rotation and expansion component causes the maximum to be greater ($\sim$260 $\pm$ 20~km~s$^{-1}$) than could be provided by expansion alone.

At location `c', ring stars are seen to be expanding towards us at $\sim$60 $\pm$ 20~km~s$^{-1}$ in agreement with the observed velocity. However, at this location velocities should be expected to change very rapidly as the expanding ring is superimposed on the receding stellar bridge (hence the mixture of blue and red points immediately above position `c').

\begin{figure}
  \centering \epsfxsize=8.5cm \epsfysize=6.5cm
  \epsffile{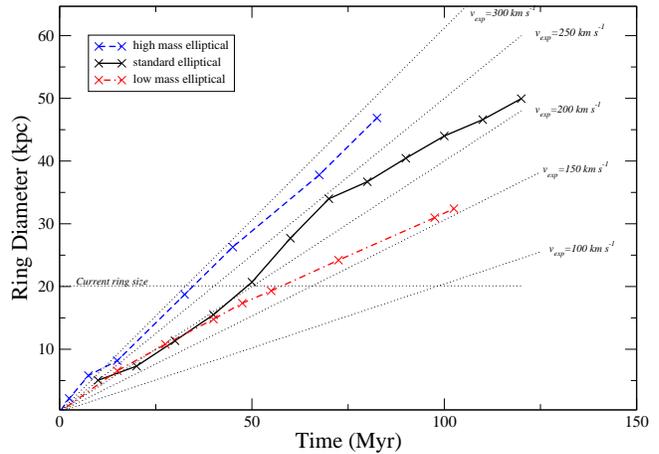}
  \caption{Evolution of the ring diameter with time and dependency on the mass of the elliptical. Thick curves indicate the results for the best-match elliptical (solid, thick, black line), low mass elliptical (dash-dot, thick, red line), and high mass elliptical (dashed, thick, blue line). The rings linearly increase their size with time in all models. The fine horizontal dash line highlights the current size of the ring. The steeper gradient of the higher mass ellipticals indicates the higher the elliptical mass, the faster the expansion of the ring. Sloping, fine dashed lines indicate how the ring diameter grows if they expand at a constant rate at the indicated velocity. Time = 0~Myr corresponds to when the ring first forms in this figure.}
\label{vexp}
\end{figure}		

\subsection{High resolution test}
We conduct a single high resolution re-run of the best-match model. In our high resolution test we double the numbers of particles of each component of both galaxies thereby halving the particle masses from those in our best-match model. This brings the total number of particles to 420,000. The high resolution model presents negligible differences from our best-match model suggesting that our best-match model has sufficient resolution to be converged.

\section{A short parameter study}
\label{paramstudy}
To conclusively prove the mechanisms by which a specific ring galaxy is formed would require exploration of a vast parameter space - despite the large number of observational constraints available for Auriga's Wheel (C11). The parameter space could potentially include the precise pre-collision morphology of both galaxies separately. Parameter space regarding the orbit of the encounter is also significant. As a result, time constraints make it unrealistic to attempt to uniformly cover all parameter space. Thus we can not claim that our scenario is the unique scenario in which Auriga's Wheel was formed. 

Instead, we conduct a parameter study, starting with our best-match model and varying a few key parameters, one at a time, to try and understand their individual impact and significance for the final ring galaxy properties. We emphasise that we repeat the best-match model  simulation with every other parameter unchanged, except varying the single parameter of interest. We choose to make our comparison between models when the ring size is 20~kpc in diameter. The parameters we vary are listed in the following:

\begin{enumerate}[(1)]
\item Total mass of the elliptical galaxy
\item Presence of bulge in (pre-collision) spiral galaxy
\item Encounter velocity
\item Impact parameter of the collision
\item Disk inclination with respect to initial velocity vector
\item Low influence parameters (spiral gas fraction, mass distribution of the elliptical)
\end{enumerate}

\begin{figure}
\begin{center}$
\begin{array}{cc}
\includegraphics[scale=0.5]{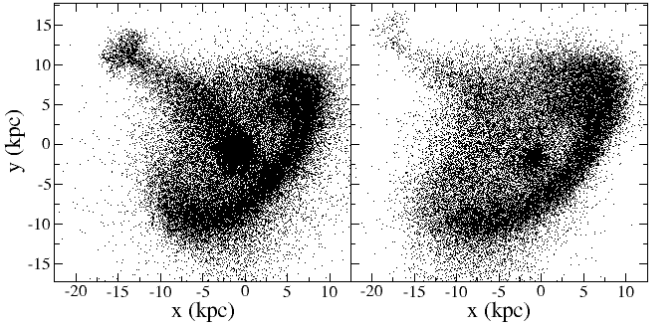} \\  
\end{array}$
\end{center}
\caption{At t = 0 Myr, the distribution of stars in the best-match model (left, includes a bulge component), and in the pure disk model (right, bulgeless). The nucleus contain approximately 6 times less stars when a bulge component is excluded from the model. Also, the thickness of the stellar bridge is reduced and it now contains $\sim$40$\%$ less stars.}
\label{nonucl}
\end{figure}

\subsection{The total mass of the elliptical galaxy}
While the stellar mass is approximately constrained based on the luminosity and colour, the dark matter content is poorly constrained. Although we have chosen our best-match total mass so that it can reproduce the observed central velocity dispersion, the post-collision observed velocity dispersion is unlikely to be a good measure of the dynamical mass as the stars are unlikely to be close to virialised within the potential well of the elliptical (for example see Fig. \ref{elplagrangerad}). Thus the quantity of dark matter or as a consequence, the total mass of the elliptical is poorly observationally constrained.

We find that the mass of the elliptical plays a key role in setting the expansion velocity of the ring. The best-match elliptical has a mass of 2.0$\times10^{11}$M$_\odot$. We model a `low mass elliptical' model which has half this mass (1.0$\times10^{11}$M$_\odot$), as if it contains no dark matter. The `high mass elliptical' is 3.5$\times10^{11}$~M$_\odot$.

In Fig. \ref{vexp}, we present the time-evolution of the diameter of the ring. To first order, the rings expand at a constant velocity between their formation and t = 0~Myr (our current view of the ring). We find that by increasing the mass of the elliptical, we increase the velocity of expansion of ring. In fact, the mass of the elliptical is found to be the key parameter controlling the velocity of ring expansion within our parameter set. At t = 0~Myr, our `low mass' elliptical expands at $\sim$100~km~s$^{-1}$, whereas our best-match model expands at $\sim$200~km~s$^{-1}$, and the `heavy mass' elliptical expands at $\sim$300~km~s$^{-1}$. Observationally, the ring expansion rate is observed to be $\sim$200~km~s$^{-1}$ assuming a circular ring inclined at 30$^\circ$ from edge-on to our line-of-sight. The model ring is also very circular, and even if we vary the disk inclination by as much as $\pm10^\circ$, the expansion velocity is must still be limited to be between 180$-220$~km~s$^{-1}$. Thus we are not free to vary the total elliptical mass significantly away from our choice in the best-match model. 

The stellar mass in the elliptical is 1.1$\times10^{11}$~M$_\odot$ from the colour and luminosity. The strong dependency of the ring expansion velocity means the total mass of the elliptical elliptical galaxy cannot differ significantly from $2.0\times10^{11}$~M$_\odot$. The dynamical mass to light ratio is therefore also constrained to be close to two. In principle, this manner of measuring the mass of the companion galaxy could provide an interesting independent measure of galaxy dark matter content. Although the dependency of expansion velocity on companion galaxy mass would first need to be understood over a far larger parameter space than we present here.   

Analytically, the ring is predicted to expand at velocity that scales with the mass of the elliptical (\citealp{Struck1990}). Comparing the expansion velocity between the `low mass' and our best-match model, we find that our models are in excellent agreement with the analytical predictions. However the `high mass' model would be analytically predicted to expand a little faster ($\sim$350~km~s$^{-1}$) than what we see in our model ($\sim$300~km~s$^{-1}$).

\begin{figure}
  \centering \epsfxsize=8.5cm \epsfysize=6.5cm
  \epsffile{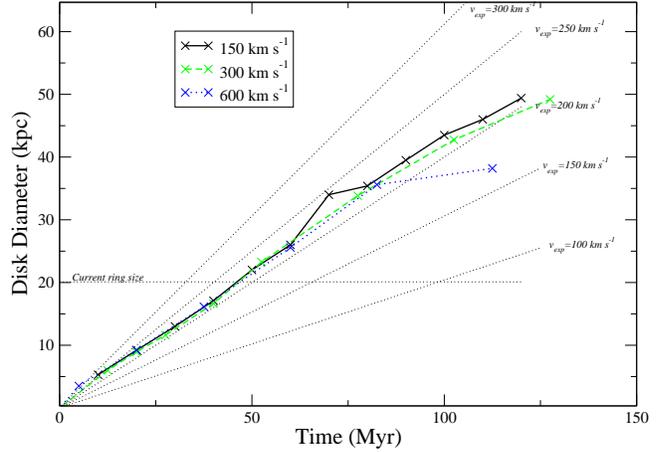}
  \caption{Evolution of ring diameter with time and dependency on the relative galaxy velocity. Thick curves indicate the results for the best-match model with initial relative velocity of 150~km~s$^{-1}$ (solid, thick, black line), an initial relative velocity of 300~km~s$^{-1}$ (dash-dot, thick, green line), and an initial relative velocity of 600~km~s$^{-1}$ (dashed, thick, blue line). The rings linearly increase their size with time in all models. The fine horizontal dash line highlights the current size of the ring. There is virtually no dependency of expansion speed on initial encounter velocity. Sloping, fine dashed lines indicate how the ring diameter grows if they expand at a constant rate at the indicated velocity. Time = 0~Myr is when the ring first forms.}
\label{vexp_velenc}
\end{figure}

\subsection{Presence of bulge in (pre-collision) spiral galaxy}
\label{nobulge}
In this test, we re-simulate our best-match model, but remove the bulge altogether from the spiral model. The original spiral is thus a pure disk model. The effect on the morphology of the stellar disk of the ring galaxy is shown in Fig. \ref{nonucl}. 

In the left panel, we show the best-match model and on the right we show the pure-disk, bulgeless model. By eye, it is clear that the mass of stars in the nucleus is significantly reduced in the bulge free model. Perhaps this should not be surprising - in Sect. \ref{nuclmorph} we noted that in the best-match model a substantial fraction of the stars in the nucleus originally came from the bulge of the pre-collision spiral. With no bulge, the mass measured with a radius of 2.5~kpc of the nucleus centre is only 2.3$\times$10$^{9}$~M$_\odot$ in stars - roughly a factor of 10 lower than observed. The colour of the nucleus in Auriga's Wheel is consistent with the observed colour of bulge stars in spirals (C11). Therefore we conclude that a bulge is required to produce the observed nucleus mass and colour. 

As in the best-match model, $\sim$50$\%$ of the spiral's gas is found within the nucleus at t = 0~Myr. If this gas is converted into stars in a starburst, this could increase the mass of the nucleus to almost 8$\times$10$^{9}~$M$_\odot$. However, unless dust obscuration results in significant reddening, this appears inconsistent with the colour of the nucleus.

Additionally the stellar bridge surface brightness is notably reduced in the bulge-free model. Recall that in the best-match model, we find that the stellar bridge consists of bulge and disk stars from the pre-collision spiral. In the absence of the bulge, we find the mass of stars in the stream is reduced to $60\%$, reducing the surface brightness of the bridge to $\sim$25~mag~arcsec$^{-2}$. In this case, the bridge would be too faint to be visible between the two galaxies. Hence the presence of the bulge is also an important factor for the strength of the stellar stream.

\subsection{Encounter velocity}
\label{encvelparam}
In the best-match model we fix the initial relative velocity of the galaxies to be 150~km~s$^{-1}$. Here we test the case of 300~km~s$^{-1}$ and 600~km~s$^{-1}$. As presented in Fig. \ref{vexp_velenc}, we find little dependency of the ring expansion speed on the initial encounter velocity.

For our best-match model, the disk reaches 20~kpc in diameter when the elliptical centre and ring galaxy nucleus have a projected separation distance of 14 $\pm$ 2~kpc - close to the observed value (see point (1)). At 300~km~s$^{-1}$, the projected separation is 22 $\pm$ 2~kpc. At 600~km~s$^{-1}$, the projected separation is 38.8 $\pm$ 2~kpc.

\begin{figure}
\begin{center}$
\begin{array}{cc}
\includegraphics[scale=0.50]{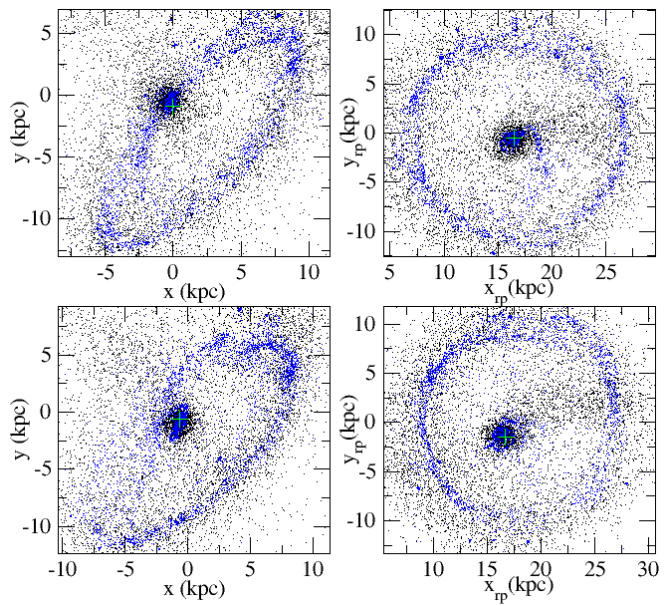} \\  
\end{array}$
\end{center}
\caption{At t = 0 Myr, the distribution of stars (black points) and gas (blue points) in the no-offset model (upper panels) and in the best-match model with offset (lower panels). Left panel shows the ring galaxy from our perspective, whereas right panel is the same ring but shown face-on (the subscript `rp' denotes that these are coordinates within the plane of the ring). The offset does not significantly deform the ring from a circular shape, but does place the nucleus off-centre within the ring when viewed the plane of the ring face-on. We highlight the centre of the nucleus with a cross symbol (green). This offset enables us to see the nucleus at approximately the correct location within the ring from our perspective (see text for more details).}
\label{nooffset}
\end{figure}

As the ring expands at equal velocities for all velocities, and we measure a modelled ring properties when it reaches a fixed diameter, it is clear that the distance separating the elliptical and the ring galaxy has become dependent on the initial relative velocity. We therefore find that we are constrained to initial relative velocities of 150~km~s$^{-1}$ in order to match the observed projected separation distance of 13~kpc.

The encounter velocity is also important for the mass of the stellar bridge. In our best-match model (150~km~s$^{-1}$), the stellar bridge mass is $\sim$5.5$\times10^{9}$M$_\odot$. At 300~km~s$^{-1}$ this mass is reduced by a factor of two. At 600~km~s$^{-1}$, the bridge does not form at all. To draw out a stream requires that a fraction of the star particles within the disk and nucleus have velocities along the axis of the interaction that are roughly of order the relative encounter velocity. In the best-match model, stars from the disk that form the bridge can clearly be seen to have left preferentially from one side of the disk. Thanks to the disk inclination, the rotation of this side provides preferential motion in the same direction as the motion of the elliptical. At 600 km~s$^{-1}$, the star particles in the disk and bulge have velocities that are simply too different from the elliptical's velocity to respond, and form the bridge. We are therefore constrained to low initial relative velocities to match both the galaxy separation and the requirement to form a stellar bridge. 
 
\subsection{A head-on collision versus a small initial offset}
In our best-match model, we arrange the galaxies so as the spiral has a small initial offset of 2.1~kpc, causing a mild offset between the centres of the galaxies at their moment of closest approach (the impact parameter is $\sim$0.75~kpc). We compare the best-match model to a no-offset model where we skip the final step (we do not include the 2.1~kpc offset), resulting in a direct head-on collision.

Results are presented in Fig. \ref{nooffset}. The right hand panels present the ring face-on, for no-offset (upper panel) versus with-offset (lower panel). We find that the offset does not significantly deform the ring away from being circular, but importantly moves the nucleus to off-centre in the ring. For clarity we have labelled the location of the centre of the nucleus with a cross symbol. The lower-right panel demonstrates that the nucleus is $\sim$2~kpc off-centre within the ring. This is important for placing the nucleus correctly within the ring when viewed from our perspective.

\begin{figure}
\begin{center}$
\begin{array}{cc}
\includegraphics[scale=0.45]{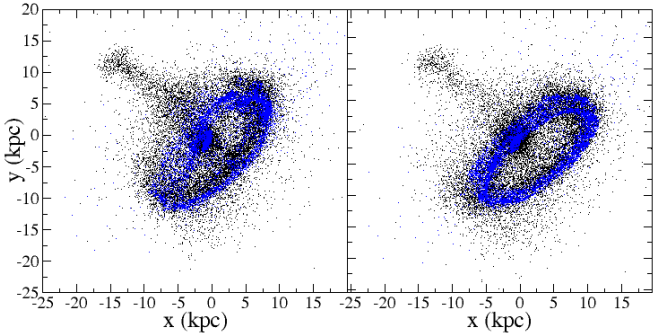} \\  
\end{array}$
\end{center}
\caption{At t=0 Myr, the distribution of stars (black points) and gas (blue points) in the best-match model with 20$^\circ$ tilt (left), and no-tilt model (right). The general morphology of the ring is approximately the same. However the nucleus is now found further behind the plane of the disk than in the best-match model, and thus appears behind the edge of the ring from our perspective. The fraction of the disk stars found in the bridge can be seen to be reduced and is actually a factor $\sim$2 less, reducing the overall mass of the stream by $\sim$35$\%$.}
\label{tilteffects}
\end{figure}

In both the best-match and `no-offset' model, the nucleus is found $\sim$$2-3$~kpc behind the plane of the ring. It is essentially slightly closer to the elliptical than the rest of the ring. Thus, when viewed from our perspective  the central nucleus appears to lie behind the ring (left panels of Fig. \ref{nooffset}). Note that the panels on the left and right correspond to the same ring, but in the right hand panel they are rotated such that we now see the ring face-on. In contrast, a slightly off-centre nucleus appears in approximately the correct location, when viewed from our perspective (lower-left panel), as seen in Auriga's Wheel. Furthermore, the off-centre collision is key in moving the nucleus away from the centre of the ring (see right panels). Therefore we find that the initial offset is important at controlling the final position of the nucleus.

\subsection{Angle of disk with respect to the initial velocity vector}
In our best-match model, the disk galaxy is mildly inclined ($20^\circ$) away from a purely face-on collision with the elliptical. We test the sensitivity of the final results to the chosen inclination angle, by conducting a no-inclination model (face-on collision).

A comparison between the disk stars and gas distribution of the best-match model (with 20$^\circ$ tilt) and the no-tilt model is shown in Fig. \ref{tilteffects}. There is little change in the general morphology of the ring. The stars and gas are distributed in a similar way. We see no indication of ring warping in either of the models although this is not unexpected at such small inclination angles (\citealp{Lynds1976}, \citealp{Ghosh2008}). More importantly however, the nucleus of the ring is now found further behind the plane of the ring - at $\sim$$5-6$~kpc (in comparison to $2-3$~kpc in the best-match model). As a result, from our perspective the nucleus now appears behind the edge of the ring in disagreement with the observed morphology. 

The tilt of the disk also has implications for the mass of the stellar bridge. We find that, in the `no-tilt' model, only half the number of disk stars are found in the bridge than in comparison to the best-match model. As discussed in Section \ref{encvelparam}, disk rotation combined with inclination, allows motion in one side of the disk to be better synchronised with the orbital motion of the elliptical causing preferential loss from this side. Without a tilt this cannot occur, and the mass of the bridge is decreased by $\sim$35$\%$. We believe that overall, a more realistic match to the observed morphology is found when we include a tilt to the disk in terms of ring shape, nucleus position and mass within the stellar stream simultaneously.

\begin{figure*}
\includegraphics[scale=0.75]{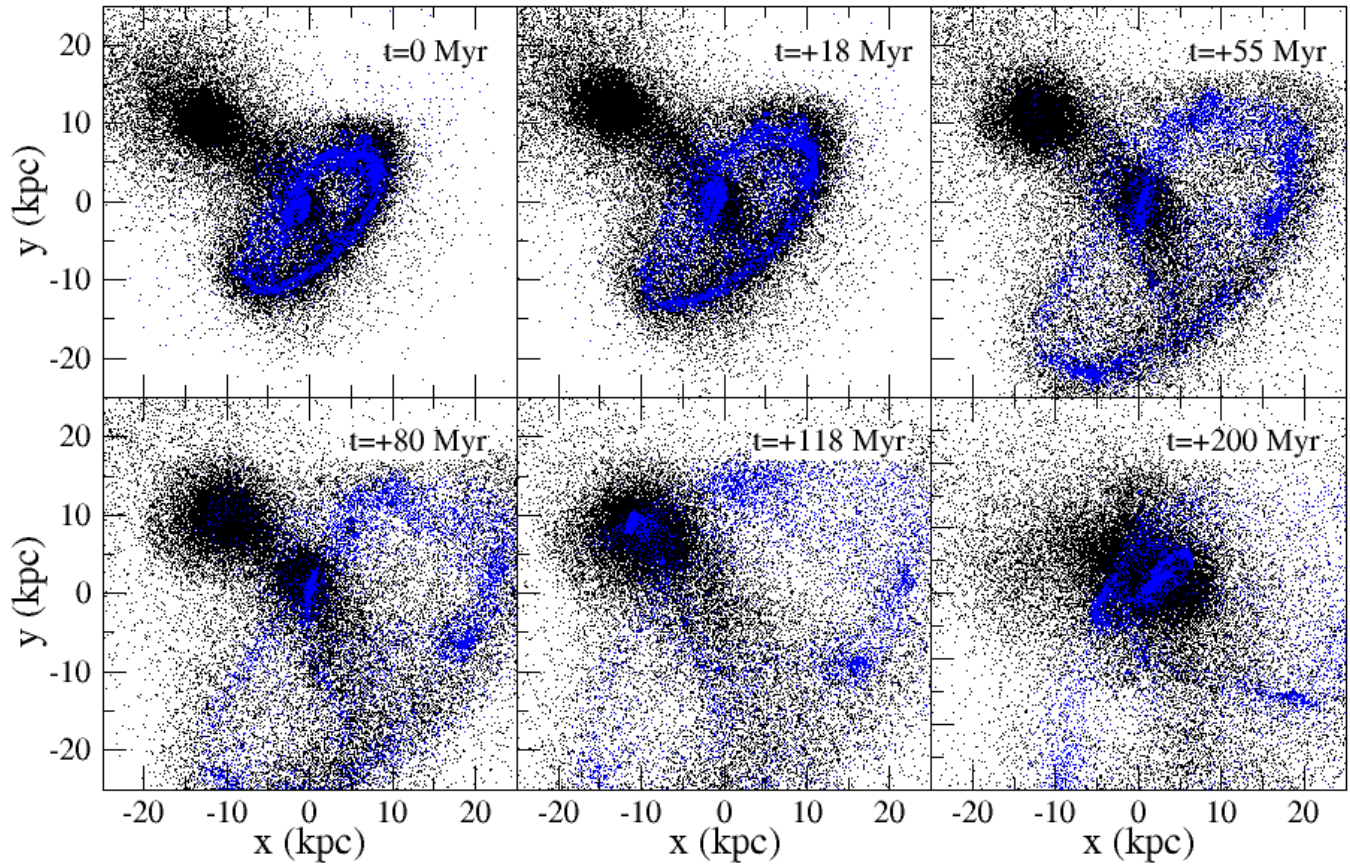}
\caption{Future evolution of the best-match model from our current view (t = 0~Myr) until t = $+200$~Myr into the future. While the ring continues to expand, the nucleus falls further behind the plane of the ring along the path of the stellar bridge. It eventually merges with the elliptical at t=$+100$~Myr. The ring ceases to expand at t$\sim$$+200$~Myr. We estimate the system will be entirely merged by t=$+400$~Myr.}
\label{futurefig}
\end{figure*}

\subsection{Low influence parameters}
\noindent
{\em{Gas fraction:}}
The gas fraction of disk in the best-match model is 20$\%$. We re-run the model but with a disk gas fraction of 10$\%$ (a value more typical of today's disk galaxies, \citealp{Gavazzi2008}). We find this is has little impact on the model. Although the gas ring is slightly less well-defined in the low gas fraction model. This is likely due to the reduced self-gravity of the ring. If high redshift disk galaxies are more gas rich than their local cousins, then high redshift CRGs could be better defined.
\newline\newline
\noindent
{\em{Mass distribution of the elliptical galaxy:}}
We test a variety of mass distributions for the elliptical galaxy. Our best-match model uses a simple, single Hernquist profile of total mass 2$\times$10$^{11}~$M$_{\odot}$, and scalelength 1.5~kpc. We also test a low concentration model (LOWC) with the scalelength doubled to 3.0~kpc. We also test models where the stellar distribution is modelled with a Hernquist profile (total mass 1$\times$10$^{11}~$M$_{\odot}$, and scalelength 1.5~kpc, identical to the best-match model), but the dark matter content is modelled with an NFW halo. We try three such models. NFWHIC has a total dark matter halo mass of 1.5$\times 10^{11}$~M$_\odot$, and a high concentration of $c$=50. NFWHIM has a more massive dark matter halo of 2.0$\times 10^{11}$~M$_\odot$, and a more typical concentration of $c$=20. NFWLOM has a light dark matter halo of 1.0$\times 10^{11}$~M$_\odot$, and concentration of $c$=20.

All models well reproduce the observed properties of Auriga's Wheel except NFWLOM. This underproduces the observed velocity of expansion of the ring by $\sim50$~km~s$^{-1}$. We note that all other elliptical models match the total mass (dark matter and stars) within a 20~kpc radius ($\sim$the current diameter of the ring) of the elliptical in the best-match model to within 15$\%$. However, the mass within 20~kpc in the NFWLOM is 30$\%$ lower, perhaps explaining the lower expansion velocity. In summary, the general properties of the system do not seem sensitive to the exact choice of mass profile for the elliptical, but rather to the total mass within the ring galaxy.

\section{A Wheelie messy future}
It is interesting to ask what the future holds for Auriga's Wheel and its elliptical companion. To attempt to answer this question, we evolve our best-match model 200~Myr forward from our current view. The resulting evolution of the stars and gas are shown in the panels of Fig. \ref{futurefig}. The upper left panel shows our the current t = 0~Myr view from our best-match model.

The expanding ring continues to grow with time. However, with increasing diameter it loses surface brightness. Meanwhile, the nucleus (which is already $\sim$$2-3$~kpc behind the plane of the ring at t = 0~Myr), falls further behind the ring along the path of the stellar bridge. Simultaneously, the elliptical decelerates significantly as a result of dynamical friction. Prior to t = 0~Myr this can be seen in Fig. \ref{relpos}. By t = $+50$~Myr, the relative velocity between the galaxies is effectively zero. As a result, the stellar bridge can remain connected to the nucleus and the elliptical. The nucleus eventually merges with the elliptical at t$\sim$$+100$~Myr, resulting in clear shells formed from star particles that were originally located within the spiral disk. We do not evolve our simulations beyond t = $+200$~Myr, as the gas in the nucleus becomes very dense following the merger between the nucleus and the elliptical, resulting in prohibitively long simulation run-time. However, it is clear that the ring expansion has decelerated and turned to collapse by t = $+200$~Myr. A complete merger of all components is inevitable. We estimate that this will occur by t$=+400$~Myr. The initial relative velocity of the encounter is simply too low for the spiral galaxy and elliptical galaxy to evade a final merger. 

\section{Conclusions}
Auriga's Wheel is a recently discovered collisional ring galaxy (C11) at redshift $z=0.11$. A close pairing between an elliptical and a ring galaxy is seen with a luminous bridge linking the galaxy pair. We study a direct collision between a spiral galaxy and elliptical galaxy as the mechanism by which Auriga's Wheel was formed using $N$-body SPH simulations.

We demonstrate that our best-match model is a reasonable match to the observations and reproduces a substantial number of key features of the observations. This includes; galaxy separation, a stellar bridge, ring diameter, ring expansion velocity and observed dynamics, a slightly off-centre nucleus of approximately the correct stellar mass, radial surface brightness and radial A4 parameter of the elliptical, and a low relative velocity.

The properties of the (precollision) spiral and elliptical galaxy in our model are not significantly different from today's galaxy. Therefore we do not expect strong differences between ring galaxies at low redshift, and those at redshift comparable to Auriga's Wheel (at $z\sim0.1$).

The best-match model provides a poor match to some observations, highlighting where the model is too idealised. The nucleus of the best-match model contains significant quantities of gas, which would likely undergo a starburst. However, the observed red colour of the nucleus in comparison to the ring apparently contradicts the model. This could indicate that our initial exponential distribution of gas within the spiral disk is incorrect. Alternatively feedback mechanisms that are not well modelled such as supernovae or AGN feedback within the nucleus could potentially act to halt a starburst within the nucleus. Excessive central inflow of gas is also a common feature in modelling of the Cartwheel galaxy (\citealp{Hernquist1993b}; \citealp{Horellou2001}; \citealp{Mapelli2008a}). 

We also note that the radially measured position-angle (and to a lesser extent, the ellipticity) of the elliptical is a poor match to the model in the inner radii. However, in the model we see that the inner radii of the elliptical are relatively unaffected by the galaxy-galaxy collision. This could suggest that the real pre-collision elliptical was mildly elliptical with a low position angle, but the outer regions were transformed during the encounter.

Assuming our best-match model is a good description of the process by which the ring galaxy formed, we can also understand the nature of the galaxies before collision, make predictions about its current nature that have not been observed, and examine its future evolution. 

\begin{enumerate}
\item Our best-match model suggests that a late-type disk galaxy suffers a direct collision with an elliptical galaxy, producing an expanding and rotating ring.
\item The total mass of the elliptical is approximately one-fifth the total mass of the spiral galaxy. The mass of the elliptical is a key parameter controlling the rate at which the ring expands. For a more massive elliptical, we produce a ring expansion that is too great compared to that observed. The velocity field of the ring is predicted, and roughly matches the values in the three locations where the ring dynamics have been measured.
\item We find that the relative velocity between the galaxies must be initially low ($\sim$150~km~s$^{-1}$) in order to reproduce the observed galaxy separation, ring diameter, and to produce the observed stellar bridge linking the elliptical to the ring galaxy. Increasing the initial relative velocity weakens the stellar bridge, and simultaneously increases the separation between the galaxies beyond that which is observed. Dynamical friction plays a significant role in reducing the orbital velocity of the elliptical as it passes through the disk galaxy.
\item We find that the elliptical galaxy suffers mild morphological transformation in its outer radii only, where as the spiral galaxy is heavily disturbed by the encounter. As a result, deriving the original properties of the spiral is difficult, as the ring galaxy bears little resemblance to its progenitor.
\item Nevertheless, our model suggests that the pre-collision spiral galaxy contained a bulge to total stellar mass (B/T) ratio $\sim$1/3. This is high but not inconsistent with B/T values in Sa-Sb type galaxies (\citealp{Weinzirl2009}). This is important as two-thirds of the stars in the nucleus of the ring galaxy had their origin in the bulge of our pre-collision spiral. If the spiral had been less bulge dominated, we would struggle to form a sufficiently massive nucleus. The colour of nucleus is consistent with an origin as bulge stars. 
\item A combination of a mild offset and a mild tilt to the disk is required to best match the observed morphology of the ring galaxy including ring appearance, size, nucleus position, and bridge mass.
\item The future of the elliptical-ring galaxy pair has largely been set by their initial low encounter velocity. When evolved into the future, we find the nucleus of the ring galaxy falls away from the plane of the ring along the stellar bridge. It eventually merges with the elliptical galaxy in a further $\sim100$~Myr. The ring continues to expand for $\sim$200~Myr then begins to recollapse. Hence we predict a total merger to occur within $\sim$400~Myr.
\end{enumerate}

A number of interesting features are highlighted in this study that may have importance for our general understanding of collisional ring galaxy formation, in particular where stellar streams are displayed. In the so-called `mushroom galaxies', a stellar stream is formed from the tidal disruption of the companion galaxy (e.g. \citealp{Wallin1994}, \citealp{Struck2003}). However our model suggests the stellar bridge in Auriga's Wheel was formed from stars originally in the disk and bulge of the parent galaxy, with a negligible contribution from the companion galaxy. For this to be possible our models critically require an initially low velocity encounter in order to draw out the stellar bridge. A mild disk inclination is additionally required to enable some orbital resonance between disk stars and the passage of the elliptical galaxy, and this boosts the contribution of disk stars to the stellar bridge.

We also highlight the heavy contribution of stars, that were bulge stars in the pre-collision disk galaxy, to the nucleus of the ring galaxy. This could contribute to the observed colour and colour gradients in ring galaxies (\citealp{Appleton1997}) as the colour and luminosity of their red inner regions may be dictated by the pre-collision bulge of the disk galaxy.

Finally we note the strong influence of dynamical friction on the future evolution of the ring-elliptical system. The rapid deceleration of the elliptical galaxy with respect to the disk galaxy is key in driving the eventual merger in our model. The effects of dynamical friction are expected to be more significant when the two galaxies have low initial collision velocities. 

One interesting example where this may be relevant is the Hoag's object (\citealp{Hoag1950}). \cite{Schweizer1987} do not favour a collisional origin for this ring galaxy as the central bulge has the same line-of-sight velocity as the ring. However \cite{Appleton1997} point out that if dynamical friction is sufficiently strong then a low relative velocity between the nucleus and ring may still be consistent with a collisional origin. In our model (see Fig. \ref{relpos}), dynamical friction plays a significant role in reducing the relative velocity between the two galaxies. 50~Myr before the collision the relative velocity is $\sim600$~km~s$^{-1}$. In the absence of dynamical friction, 50~Myr after the collision the relative velocity would again be $\sim$600~km~s$^{-1}$. In fact, due to the substantial role played by dynamical friction, 50~Myr after the collision the relative velocity has been reduced to zero.

\section*{Acknowledgements}
RS acknowledges support from a COMITE MIXTO grant, R.R.L. acknowledges support from the Chilean {\sl Centro de Astrof\'\i sica} FONDAP No. 15010003., B.C.C was financed through a ESO Fellowship and Alexander von Humboldt Foundation Fellowship, MF acknowledges support from FONDECYT grant 1095092.

\bibliography{bibfile}

\bsp

\label{lastpage}

\end{document}